\documentclass[cernpreprint, atlasdraft=false, texlive=2023, UKenglish, texmf, orcidlogo]{atlasdoc}
 
\usepackage{atlaspackage}
\usepackage{atlasbiblatex}
 
\usepackage{atlasphysics}

\usepackage{multirow}
 
\graphicspath{{logos/}{figures/}}
 
\usepackage{ANA-TOPQ-2021-08-PAPER-defs}

\AtlasTitle{Observation of four-top-quark production in the multilepton final state with the ATLAS detector}
 
\AtlasVersion{3.1}
 
\AtlasAbstract{
This paper presents the observation of four-top-quark ($t\bar{t}t\bar{t}$) production in proton-proton collisions at the LHC.
The analysis is performed using an integrated luminosity of 140~fb$^{-1}$ at
a centre-of-mass energy of 13 TeV collected using the ATLAS detector.
Events containing two leptons with the same electric charge or at least three leptons (electrons or muons) are selected.
Event kinematics are used to separate signal from background through a multivariate discriminant, and dedicated control regions are used to constrain the dominant backgrounds.
The observed (expected) significance of the measured $t\bar{t}t\bar{t}$ signal with respect to the standard model (SM) background-only hypothesis is 6.1 (4.3) standard deviations. The $t\bar{t}t\bar{t}$ production cross section is measured to be $22.5^{+6.6}_{-5.5}$~fb, consistent with the SM prediction of $12.0 \pm 2.4$ fb within 1.8 standard deviations. Data are also used
to set limits on the three-top-quark production cross section,
being an irreducible background not measured previously,
and to constrain the top-Higgs Yukawa coupling and effective field theory operator coefficients that affect $t\bar{t}t\bar{t}$ production.
}

\AtlasRefCode{TOPQ-2021-08}
 
\PreprintIdNumber{CERN-EP-2023-055}

\AtlasJournalRef{Eur. Phys. J. C 83 (2023) 496}
\AtlasDOI{10.1140/epjc/s10052-023-11573-0}

\hypersetup{pdftitle={ATLAS document},pdfauthor={The ATLAS Collaboration}}
 
\begin{document}
 
\maketitle
\tableofcontents

\section{Introduction}
 
The top quark is the heaviest elementary particle in the Standard Model (SM) and has a strong connection to the SM Higgs boson as well as potentially new particles in various theories beyond the SM (BSM). It is therefore relevant to study rare processes involving the top quark, such as the production of four top quarks (\tttt), which is predicted by the SM but has not been observed yet.
The \tttt\ cross section could be enhanced in many BSM models, including gluino pair production in supersymmetric theories~\cite{Nilles:1983ge,Farrar:1978xj}, scalar-gluon pair production~\cite{Plehn:2008ae,Calvet:2012rk},
the associated production of a heavy scalar or pseudoscalar boson with a top-quark pair in two-Higgs-doublet models~\cite{Dicus:1994bm,Craig:2015jba,Craig:2016ygr},
or in top-quark-compositeness models~\cite{PhysRevD.78.074026}.
Additionally, the \tttt cross section is sensitive to the strength of the top-quark Yukawa coupling, and its charge conjugation and parity ($CP$) properties~\cite{Cao:2017,Cao:2019ygh}. It is also sensitive to various four-fermion interactions~\cite{Degrande:2010kt,Zhang:2017mls,Banelli:2020iau, Aoude:2022deh} and the Higgs oblique parameter~\cite{Englert:2019zmt} in the context of an effective field theory (EFT) framework.
 
\begin{figure}[!htb]
\centering
\includegraphics[width=0.4\textwidth]{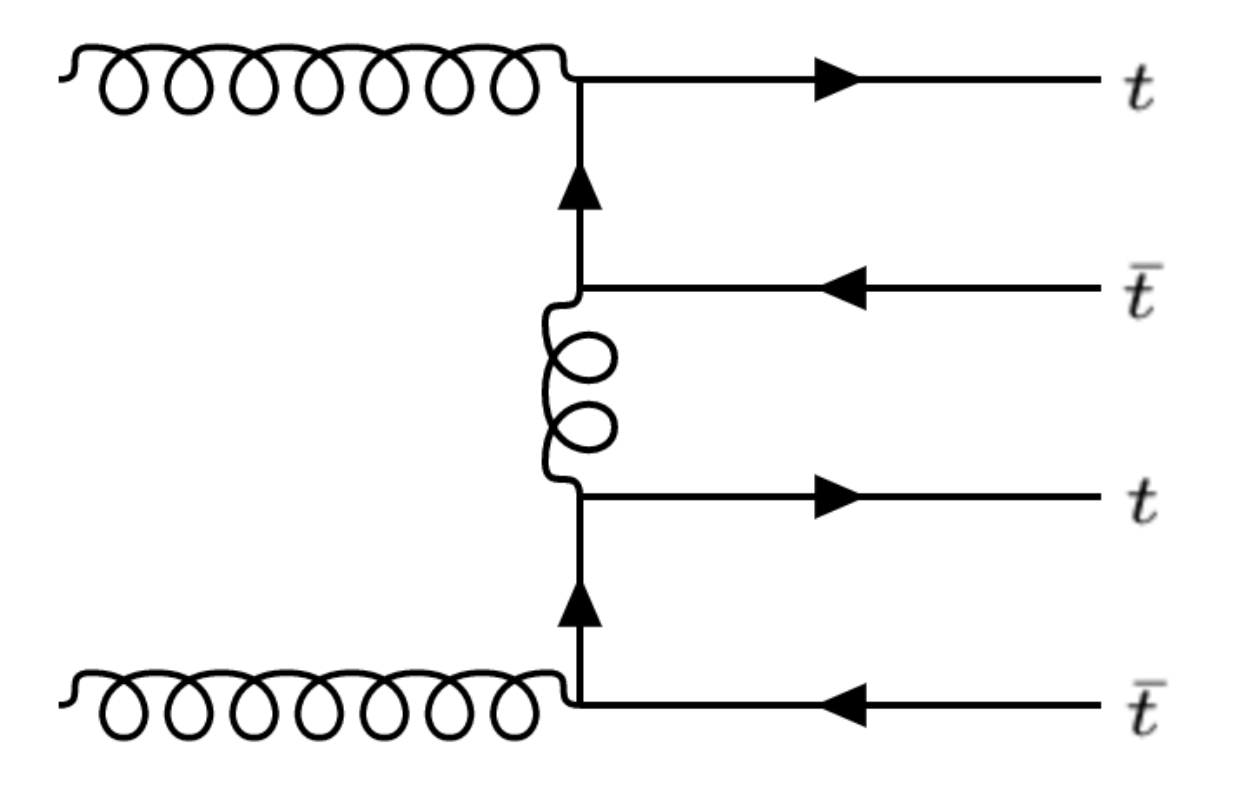}
\includegraphics[width=0.4\textwidth]{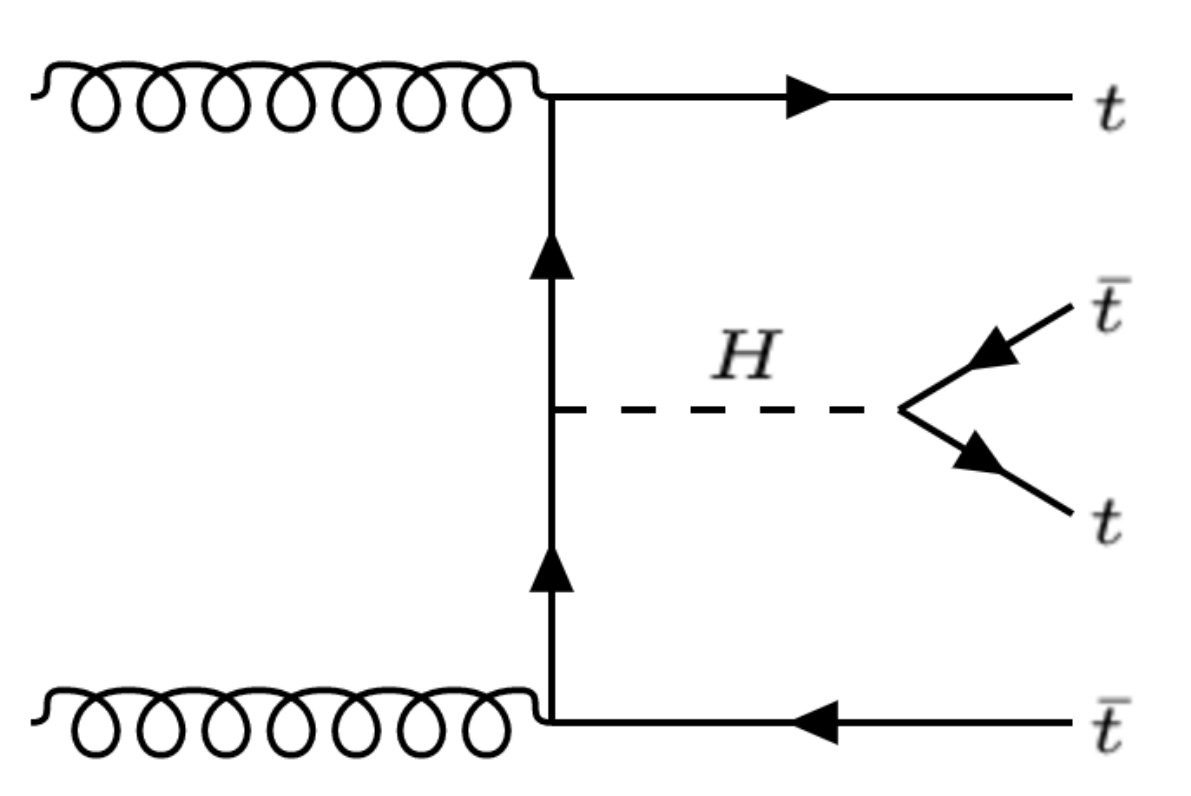}
\caption{Illustrative tree-level Feynman diagrams for the SM \tttt signal. The mediator connecting two top quarks can be a gluon, a neutral electroweak gauge boson ($\gamma$ or $Z$), or a Higgs boson.}
\label{fig:feynman}
\end{figure}

Within the SM, the predicted cross section of \tttt\ production in proton-proton ($pp$) collisions at a centre-of-mass energy $\sqrt{s}=13$~TeV is $\sigma_{\tttt}=12.0$~fb at next-to-leading order (NLO) in QCD including electroweak corrections, with a relative scale uncertainty of ±20\%~\cite{Frederix:2017wme, Bevilacqua:2012em, Jezo:2021smh}. This prediction is used as a reference throughout this paper, as for the previous publications~\cite{TOPQ-2018-05,ATLAS:2021kqb}.
Including threshold resummation at next-to-leading logarithmic accuracy increases the total production cross section by approximately 12\% and reduces the scale uncertainty significantly, leading to $\sigma_{\tttt}=13.4^{+1.0}_{-1.8}$~fb~\cite{vanBeekveld:2022hty}.
Illustrative Feynman diagrams for SM \tttt\ production are shown in Figure~\ref{fig:feynman}.
Other rare multi-top-quark production processes which haven't been observed yet, such as three-top-quark production (\ttt)\footnote{In the following, $t\bar{t}t$ is used to denote both $t\bar{t}t$ and its charge conjugate, $\bar{t}t\bar{t}$. Similar notation is used for other processes, as appropriate.},
are expected to have cross sections of $O$(1~fb) in the SM, an order of magnitude smaller than \tttt\ production. They can also be sensitive to BSM effects.
 
The \tttt\ process results in various final states depending on the top quark decays.
These final states are classified based on the number of electrons or muons produced in the semileptonic top quark decays, including those originating from subsequent $\tau$ leptonic decays. This paper focuses on two types of events: those with exactly two same-charge isolated leptons (2LSS) and those with at least three isolated leptons (3L).
In the SM, 7\% and 5\% of the produced four top-quark events result in 2LSS and 3L final states, respectively.
While these channels, referred to as 2LSS/3L,
are relatively rare final states, they are advantageous due to low levels of background.
The \tttt\ topology is also characterised by a high light-jet and $b$-jet multiplicity and a large rest mass of the event, amounting to about $700~\gev$.
 
The ATLAS and CMS experiments have reported evidence for \tttt\ production in 13 TeV $pp$ collisions at the LHC.
The latest ATLAS result combines two analyses using 139~\ifb\ at $\sqrt{s}=13$~TeV: one in the 2LSS/3L channel~\cite{TOPQ-2018-05} and the other in the channel comprising events with one lepton or two leptons with opposite electric charge.
This combination results in a measured cross-section of
$24^{+7}_{-6}$~fb, corresponding to an observed (expected) signal significance of 4.7 (2.6) standard deviations over the background-only predictions~\cite{ATLAS:2021kqb}.
The latest CMS result is from a combination of several measurements using 138~\ifb\ at $\sqrt{s}=13$~TeV, in the channels with zero, one and two electrons or muons with opposite-sign electric charges and the 2LSS/3L channel, yielding an observed (expected) significance of 4.0 (3.2) standard deviations~\cite{CMS:2023zdh}.
The \tttt cross section measured by the CMS collaboration is $17 \pm 5$~fb.
 
This paper presents a re-analysis of the 140~\ifb\ data set at $\sqrt{s}=13$~TeV in the 2LSS/3L channel with the ATLAS detector and supersedes the result of Ref.~\cite{TOPQ-2018-05}. Compared to the previous result that showed evidence for \tttt\ production~\cite{TOPQ-2018-05}, this new measurement brings several improvements: an optimised selection with lower cuts on the leptons' and jets' transverse momenta; improved $b$-jet identification; a new data-driven estimation of the \ttWj background, one of the main backgrounds in this channel; a revised set of systematic uncertainties; an improved treatment of the \ttt background and a more powerful multivariate discriminant to separate the signal from background. This paper also presents limits under several BSM scenarios,
such as limits on the three-top-quark production cross section, on the top-Higgs Yukawa coupling, and on EFT operators corresponding to heavy-flavour four-fermion interactions and the Higgs oblique parameter.


\section{The ATLAS detector}
\newcommand{\AtlasCoordFootnote}{
ATLAS uses a right-handed coordinate system with its origin at the nominal interaction point (IP)
in the centre of the detector and the \(z\)-axis along the beam pipe.
The \(x\)-axis points from the IP to the centre of the LHC ring,
and the \(y\)-axis points upwards.
Cylindrical coordinates \((r,\phi)\) are used in the transverse plane,
\(\phi\) being the azimuthal angle around the \(z\)-axis.
The pseudorapidity is defined in terms of the polar angle \(\theta\) as \(\eta = -\ln \tan(\theta/2)\).
Angular distance is measured in units of \(\Delta R \equiv \sqrt{(\Delta\eta)^{2} + (\Delta\phi)^{2}}\).}
 
The ATLAS detector~\cite{PERF-2007-01} at the LHC covers nearly the entire solid angle around the collision point.\footnote{\AtlasCoordFootnote}
It consists of an inner tracking detector surrounded by a thin superconducting solenoid, electromagnetic and hadron calorimeters,
and a muon spectrometer incorporating three large superconducting air-core toroidal magnets.
 
The inner-detector system (ID) is immersed in a \qty{2}{\tesla} axial magnetic field
and provides charged-particle tracking in the range \(|\eta| < 2.5\).
The high-granularity silicon pixel detector covers the vertex region and typically provides four measurements per track,
the first hit normally being in the insertable B-layer (IBL) installed before Run~2~\cite{ATLAS-TDR-19,PIX-2018-001}.
It is followed by the silicon microstrip tracker (SCT), which usually provides eight measurements per track.
These silicon detectors are complemented by the transition radiation tracker (TRT),
which enables radially extended track reconstruction up to \(|\eta| = 2.0\).
The TRT also provides electron identification information
based on the fraction of hits (typically 30 in total) above a higher energy-deposit threshold corresponding to transition radiation.
 
The calorimeter system covers the pseudorapidity range \(|\eta| < 4.9\).
Within the region \(|\eta|< 3.2\), electromagnetic calorimetry is provided by barrel and
endcap high-granularity lead/liquid-argon (LAr) calorimeters,
with an additional thin LAr presampler covering \(|\eta| < 1.8\)
to correct for energy loss in material upstream of the calorimeters.
Hadron calorimetry is provided by the steel/scintillator-tile calorimeter,
segmented into three barrel structures within \(|\eta| < 1.7\), and two copper/LAr hadron endcap calorimeters.
The solid angle coverage is completed with forward copper/LAr and tungsten/LAr calorimeter modules
optimised for electromagnetic and hadronic energy measurements respectively.
 
The muon spectrometer (MS) comprises separate trigger and
high-precision tracking chambers measuring the deflection of muons in a magnetic field generated by the superconducting air-core toroidal magnets.
The field integral of the toroids ranges between \num{2.0} and \qty{6.0}{\tesla\metre}
across most of the detector.
Three layers of precision chambers, each consisting of layers of monitored drift tubes, covers the region \(|\eta| < 2.7\),
complemented by cathode-strip chambers in the forward region, where the background is highest.
The muon trigger system covers the range \(|\eta| < 2.4\) with resistive-plate chambers in the barrel, and thin-gap chambers in the endcap regions.
 
Interesting events are selected by the first-level trigger system implemented in custom hardware,
followed by selections made by algorithms implemented in software in the high-level trigger~\cite{TRIG-2016-01}.
The first-level trigger accepts events from the \qty{40}{\MHz} bunch crossings at a rate below \qty{100}{\kHz},
which the high-level trigger further reduces in order to record events to disk at about \qty{1}{\kHz}.
 
An extensive software suite~\cite{ATL-SOFT-PUB-2021-001} is used in data simulation, in the reconstruction
and analysis of real and simulated data, in detector operations, and in the trigger and data acquisition
systems of the experiment.


\section{Data and Monte Carlo simulation samples}
\label{sec:mcsamples}
 
This analysis was performed using the $pp$ collision data collected by the ATLAS detector between 2015 and 2018 at $\sqrt{s} = \SI{13}{\TeV}$.
After the application of data-quality requirements~\cite{DAPR-2018-01},
the data set corresponds to an integrated luminosity of 140~\ifb.
 
Samples of simulated, or Monte Carlo (MC), events were produced to model the different signal and background processes. Additional samples were produced to estimate the modelling uncertainties for each process. The configurations used to produce the samples are summarised in Table~\ref{tab:mc_config}.
The effects of the additional \pp\ collisions in the same or a nearby bunch crossing (pile-up) were modelled by overlaying minimum bias events simulated using \pythia.186~\cite{Sjostrand:2014zea} with the A3 set of tuned parameters (tune)~\cite{ATL-PHYS-PUB-2016-017} on events from hard-scatter processes.
The MC events were weighted to reproduce the distribution of the average number of interactions per bunch crossing observed in the data.
All samples generated with \powhegbox~\cite{Nason:2004rx,Frixione:2007nw,Frixione:2007vw,Alioli:2010xd} and \mgamc~\cite{Frederix:2012ps} were interfaced to \pythia to simulate the parton shower, fragmentation, and underlying event with the A14 tune~\cite{ATL-PHYS-PUB-2014-021} and the \NNPDF[2.3lo] distribution function (PDF) set.
Samples using \pythia and \herwigseven have heavy-flavour hadron decays modelled by EvtGen v1.2.0 or EvtGen v1.6.0~\cite{Lange:2001uf}.
All samples include leading-logarithm photon emission, either modelled by the parton shower generator or by \textsc{PHOTOS}~\cite{Golonka:2005pn}. The masses of the top quark, \mtop, and of the SM Higgs boson, $m_H$, are set to $172.5~\gev$ and $125~\gev$, respectively.
The signal and background samples were processed through the simulation of the ATLAS detector geometry and response
using \textsc{Geant4}~\cite{Agostinelli:2002hh}, and then reconstructed using the same software as for collider data. Some of the alternative samples used to evaluate systematic uncertainties were instead processed through a fast detector simulation making use of parameterised showers in the calorimeters~\cite{ATL-PHYS-PUB-2010-013}.
 
\begin{table}
\renewcommand*{\arraystretch}{1.4}
\caption{Summary of the generation setups used for the signal and background simulation.
The samples used to estimate the systematic uncertainties are indicated in parentheses.
$V$ refers to production of an electroweak boson ($W$ or $Z/\gamma^*$).
The matrix element (ME) order refers to the order in QCD of the perturbative calculation.
The parton distribution function (PDF) used for the ME is shown in the table.
Tune refers to the underlying-event tune of the parton shower (PS) generator.  \MEPSatNLO and FxFx refer to
the methods used  to
match the ME to the PS in \sherpa~\cite{Cascioli:2011va,Gleisberg:2008fv,Schumann:2007mg,Hoeche:2012yf} and in \mgamc~\cite{Frederix_2012}, respectively.}
\label{tab:mc_config}
\begin{center}{\small
\setlength\tabcolsep{4.0pt}
\begin{tabular}{llllll}
\hline\hline
Process & Generator & ME order  & PDF & Parton shower & Tune  \\
\hline
\tttt & \mgamc & NLO &  \NNPDF[3.1nlo]  & \pythia & A14\\
& (\mgamc) & (NLO) & (MMHT2014 LO) & (\herwigseven) & (H7UE) \\
&(\sherpa~2.2.11) & (NLO) & (\NNPDF[3.0nnlo]) & (\sherpa) & (\sherpa) \\
& (\mg) & (LO) &  (MMHT2014 LO)& (\pythia) & (A14) \vspace{4pt}\\
\tttt $\kappa_t$ & \mg & LO & MMHT2014 LO & \pythia & A14 \vspace{4pt}\\
\tttt EFT    & \mg & LO & MMHT2014 LO & \pythia & A14 \\
\hline
\ttt & \mg &  LO & \NNPDF[2.3lo] & \pythia & A14 \vspace{4pt}\\
\ttW  & \sherpa~2.2.10 & \MEPSatNLO &  \NNPDF[3.0nnlo] & \sherpa & \sherpa \\
& (\mgamc) & (FxFx)& (\NNPDF[2.3lo]) & (\pythia) & (A14) \vspace{4pt}\\
\ttW EW & \sherpa~2.2.10 & LO & \NNPDF[3.0nnlo] & \sherpa & \sherpa \\
& (\mg) & (LO)&  (\NNPDF[2.3lo] ) & (\pythia)  & (A14) \vspace{4pt}\\
\ttZ & \mgamc & NLO &  \NNPDF[3.0nlo] & \pythia & A14 \\
&(\sherpa~2.2.11 ) & (\MEPSatNLO)  & (\NNPDF[3.0nnlo]) & (\sherpa) & (\sherpa) \vspace{4pt}\\
\ttH  & \powhegbox & NLO  & \NNPDF[3.0nlo] &  \pythia & A14\\
& (\powhegbox) & (NLO) &  (\NNPDF[3.0nlo]) & (\herwigseven)  & (H7UE)\\
& (\mgamc) &  (NLO) &  (\NNPDF[3.0nlo]) & (\pythia) & (A14) \vspace{4pt}\\
\tt & \powhegbox & NLO &   \NNPDF[3.0nlo] & \pythia  & A14\\
Single-top &  \powhegbox & NLO &  \NNPDF[3.0nlo] & \pythia  & A14\\
\tWZ & \mgamc & NLO & \NNPDF[3.0nlo]  & \pythia & A14 \vspace{4pt}\\
\tZq & \mg &  LO  & \NNPDF[2.3lo] & \pythia & A14 \vspace{4pt}\\
$t\bar{t}VV$ & \mg &  LO  & \NNPDF[2.3lo] & \pythia & A14 \vspace{4pt}\\
$V$ & \sherpa~2.2.1 & \MEPSatNLO & \NNPDF[3.0nnlo] & \sherpa & \sherpa  \vspace{4pt}\\
$VV$,$VVV$ &  \sherpa~2.2.2 & \MEPSatNLO &  \NNPDF[3.0nnlo] & \sherpa & \sherpa  \vspace{4pt}\\
$VH$ & \pythia & LO  &\NNPDF[2.3lo] & \pythia & A14\\
\hline\hline
\end{tabular}
}
\end{center}
\end{table}

The nominal sample used to model the SM \tttt signal was generated using the \mgamc{v2.6.2} \cite{Alwall:2014hca} generator, which provides matrix elements at NLO in QCD. The renormalisation ($\mu_R$) and factorisation ($\mu_F$) scales are set to $\mu_R = \mu_F = m_{\texttt{T}}/4$,
where $m_{\texttt{T}}$ is the scalar sum of the transverse masses
$\sqrt{p_{\texttt{T}}^{2} + m^{2}}$ of the particles generated from the matrix element calculation, following Ref.~\cite{Frederix:2017wme}.
An alternative \tttt sample generated with the same \mgamc set-up, but interfaced to \herwigseven.04~\cite{Bahr:2008pv,Bellm:2015jjp} with the H7UE tune~\cite{Bellm:2015jjp}, is used to evaluate uncertainties due to the choice of parton shower and hadronisation model.
Another \tttt\ sample was produced at NLO using the \sherpa~2.2.11~\cite{Bothmann:2019yzt} generator to account for the modelling uncertainty from an alternative generator choice. The scales for this sample are set to $\mu_R=\mu_F=\hat{H}_\textrm{T}/2$,
where $\hat{H}_\textrm{T}$ is defined as the scalar sum of the transverse momenta of all final state particles.
To mitigate the effect of the large fraction of events with negative weights present in the nominal sample that would be
detrimental to the training of the multivariate discriminant used to separate signal from background, an additional sample with settings similar to the nominal ones was generated using leading order (LO) matrix elements (MEs).
The \tttt MC samples with different BSM effects such as non-SM top Yukawa coupling and various EFT operators, including the four-fermion and the Higgs oblique operators, were generated at LO with \mg. They are referred to as ``\tttt $\kappa_t$'' and ``\tttt EFT'' in Table~\ref{tab:mc_config}.
Dedicated UFO models implemented in \mg\ with FeynRules~\cite{Alloul:2013bka,Degrande:2011ua} are used to simulate these BSM effects.
The alternative Yukawa coupling scenarios were simulated using the Higgs characterisation model~\cite{Artoisenet:2013puc}.
The effects of the four-fermion and Higgs oblique operators are simulated using the SMEFT@NLO model~\cite{smeftatnlo} and the dedicated UFO model from Ref.~\cite{Englert:2019zmt}, respectively.
 
The \ttW process was simulated using the \sherpa~2.2.10~\cite{Bothmann:2019yzt,Gleisberg:2008ta} generator. The MEs were calculated for up to one additional parton at NLO and up to two partons at LO using \textsc{Comix}~\cite{Gleisberg:2008fv}
and \OPENLOOPS~\cite{Cascioli:2011va}, and merged with the \Sherpa
parton shower~\cite{Schumann:2007mg} using the \MEPSatNLO
prescription~\cite{Hoeche:2012yf} with a merging scale of $30~\gev$.
The choice of scales is $\mu_R = \mu_F =
m_\textrm{T}$/2.
In addition to the nominal prediction at NLO in QCD, higher-order corrections related to electroweak (EW)
contributions are also included. Event-by-event correction factors are applied that provide virtual NLO EW corrections to O($\alpha^{2}\alphas^{2}$) and LO corrections to O($\alpha^3$) ~\cite{Kallweit:2015dum,Bothmann:2019yzt,Gutschow:2018tuk}.
An independent \Sherpa~2.2.10 sample was produced at LO to account for the sub-leading EW corrections to O($\alpha^{3}\alphas$)~\cite{Frederix:2017wme}.
The NLO QCD and NLO EW contributions from \Sherpa are combined following the method of  Ref.~\cite{Frederix:2021agh}.
The \ttW samples are normalised using a cross section of 722~fb computed at NLO including the hard non-logarithmically enhanced radiation at NLO in QCD~\cite{Frederix:2021agh}. This normalisation is used to determine the contribution of the \ttWj background
before additional data-driven corrections are derived.
The impact of the systematic uncertainty associated with the choice of generator is evaluated using an alternative \ttW sample generated with up to one additional parton in the final state at NLO accuracy in QCD
using \mgamc. In this sample, the different jet multiplicities were merged using the FxFx NLO matrix-element and parton-shower merging prescription~\cite{Frederix_2012} with a merging scale of $30~\gev$. A separate \mg \ttWj sample including  only the sub-leading EW corrections to O($\alpha^{3}\alphas$) is also included.
 
The \ttZ sample was generated at NLO, including off-shell $Z$ contributions with $m(\ell^+\ell^-) > 1~\gev$ and is normalised to the calculation of Ref.~\cite{deFlorian:2016spz}, extrapolated to take into account off-shell effects down to $m(\ell^+\ell^-) = 1~\gev$.
The \ttH\ process was generated at NLO QCD using the five-flavour scheme with the \hdamp\ parameter\footnote{The \hdamp\ parameter controls the transverse momentum \pt\ of the first additional emission beyond the leading-order Feynman diagram in the parton shower and therefore regulates the high-\pt\ emission against which the \tt\ system recoils.}
set to $0.75\times(2\mtop+m_H)$. It is normalised to the cross section computed at NLO QCD plus NLO EW ~\cite{Frixione:2015zaa,deFlorian:2016spz}. The production of \tt\ and single-top-quark events were modelled at NLO QCD using \powhegbox. For the \tt\ sample, the \hdamp parameter is set to 1.5~\mtop~\cite{ATL-PHYS-PUB-2016-020}. The \tt\ and single-top-quark simulated samples are normalised to the cross sections
calculated at next-to-next-to-leading order (NNLO) QCD including the
resummation of next-to-next-to-leading logarithmic (NNLL) soft-gluon terms~\cite{Czakon:2011xx,Kidonakis:2010ux,Kidonakis:2010tc,Kidonakis:2011wy}. The overlap between the \tt\ and the single top $tW$ final
states is removed using the diagram-removal scheme~\cite{Frixione:2008yi}.
The \ttt sample was generated at LO QCD using the five-flavour scheme, and includes \tttW and \tttq processes. The \tttW and \tttq samples are normalised to a cross section of 1.02~fb and 0.65~fb, respectively, computed at NLO using \mgamc, with a central scale choice of $\hat{H}_{\texttt{T}}/2$ and the NNPDF23 PDF set.
Rare and small background contributions including \tWZ, \tZq, \ttVV, \ttHH, \ttWH, $V$, $VV$, $VVV$ and $VH$ are normalised using their NLO theoretical cross sections.


\section{Objects and event selection}
 
Events were selected using single-lepton or dilepton triggers with variable electron and
muon transverse momentum (\pT) thresholds, and various identification and isolation criteria
depending on the lepton flavour and the data-taking period~\cite{TRIG-2018-01,TRIG-2018-05,TRIG-2019-04_sp,TRIG-2019-03}.
Events are required to have at least one vertex with at least two associated ID tracks with $\pt > 0.5$~\GeV.
In each event, the primary vertex is defined as the reconstructed vertex having the highest scalar
sum of squared \pt of associated tracks~\cite{ATL-PHYS-PUB-2015-026} and is consistent with the average beam-spot position.
 
Electron candidates are reconstructed from energy deposits in the EM
calorimeter associated with ID tracks~\cite{EGAM-2018-01} and are required to consist of
a calorimeter energy cluster with a pseudorapidity of  $|\eta_{\text{cluster}}|<2.47$,
excluding candidates in the transition region between the barrel and the endcap
calorimeters (${|\eta_{\text{cluster}}|\not\in[1.37,1.52]}$). Electrons are
required to meet the `tight' likelihood-based identification criterion~\cite{EGAM-2018-01}.
Muon candidates are reconstructed by combining tracks in
the ID with tracks in the muon spectrometer~\cite{CERN-EP-2020-199}. They are required to have $|\eta|<2.5$
and meet the `medium' cut-based identification criterion.
 
The electron and muon candidates are required to have $\pT>$~15~\GeV\ and meet the
tight working point of the prompt-lepton isolation discriminant~\cite{HIGG-2017-02}, trained to
separate prompt and non-prompt leptons.
The transverse impact parameter divided by its estimated
uncertainty, ${|\dzero|/\sigma(\dzero)}$, is required to be lower
than five (three) for electron (muon) candidates. The longitudinal impact
parameter must satisfy ${|\zzsth|<0.5}$~mm for both lepton flavours.
To suppress the background arising from  incorrect charge assignment, an additional
requirement is imposed on electrons based on the score of a boosted decision tree (BDT) discriminant
that uses their calorimeter energy clusters and track properties~\cite{EGAM-2018-01}.
Electrons and muons passing these criteria are referred to as `tight' in the following.
`Loose' electrons are required to pass the same set of requirements as tight electrons except that they satisfy
the `medium' likelihood-based identification and loose isolation instead.
Similarly, for `loose' muons, the isolation requirement is relaxed.
 
Jets are reconstructed using a particle flow algorithm~\cite{PERF-2015-09,JETM-2018-05_sp} by combining
measurements from both the ID and the calorimeter. The \antikt algorithm~\cite{Cacciari:2008gp,Fastjet}
with a radius parameter of \(R=0.4\) is used. Jets are required to have $\pT>\SI{20}{\GeV}$
and $|\eta|<2.5$ and are calibrated as described in Ref.~\cite{JETM-2018-05_sp}.
To reduce the effect from pile-up,
the jet-vertex tagger~\cite{ATLAS-CONF-2014-018}, which identifies jets originating from the selected primary vertex, is applied to jets with $\pt < 60~\GeV $ and $|\eta|<2.4$.
 
Jets containing $b$-hadrons are identified ($b$-tagged) via the \texttt{DL1r} algorithm~\cite{r21_dl1r,ATL-PHYS-PUB-2017-013}.
This algorithm uses deep-learning neural networks exploiting the distinct features of $b$-hadrons in terms of track impact parameters and displaced vertices reconstructed in the ID.
A jet is considered $b$-tagged if it passes the operating point corresponding to 77\% average
efficiency for $b$-quark jets in simulated $t\bar{t}$ events, with rejection factors
against light-quark/gluon jets and $c$-quark jets of 192 and 6, respectively.
Three additional operating points are defined with average $b$-tagging efficiencies of 85\%, 70\% and 60\%.
To fully exploit the $b$-tagging information of an event, each jet is assigned a pseudo-continuous $b$-tagging (PCBT) score that defines if a jet passes a given operating point but fails the adjacent tighter one.
A score of two, three, four or five is assigned to a jet passing the 85\%, 77\%, 70\% or 60\% operating point.
If a jet does not pass any operating point, a score of one is assigned.
 
The missing transverse momentum in the event, whose magnitude is denoted in the
following by \met, is defined as the negative vector sum of the \pT\ of the reconstructed
and calibrated objects in the event~\cite{PERF-2016-07}. This sum also includes the momenta of the ID tracks
that are matched to the primary vertex but are not associated with any other reconstructed objects.
 
A sequential overlap removal procedure described in Ref.~\cite{TOPQ-2018-05} and based on loose leptons is applied,
such that the same calorimeter energy deposit or the same track being reconstructed is not used in two different objects.
 
The event selection closely follows that of Ref.~\cite{TOPQ-2018-05}.
Each event must have at least one reconstructed lepton that matches a lepton that fired the trigger. The
highest \pt lepton is required to have $\pt > 28$~\GeV\ to be always above the trigger threshold.
The events are required to have one same-sign tight lepton
pair or at least three tight leptons, and at least one $b$-tagged jet. Each event must have at least one reconstructed lepton that matches a lepton that fired the trigger. Events with two same-sign electrons are required to have invariant mass satisfying $m_{ee}>15~\GeV$ and $|m_{ee}-91~\GeV|>10~\GeV$, to reduce the charge mis-assignment background originating from low-mass resonances and $Z$ boson decays. In events with at least three leptons, all opposite-sign same-flavour lepton pairs are required to satisfy $|\mll-91~\GeV|>10~\GeV$ to reduce the contamination from $Z$ boson decay.
 
Events in the signal region (SR) are required to contain at least six jets, two of which are $b$-tagged.
The scalar sum of the transverse momentum of the selected leptons and jets in the event (\HT) is required to be above $500\,\GeV$.
This requirement is motivated by the high light-flavour jet and $b$-jet
multiplicity as well as the large overall event activity characteristic of the \tttt process.
 
The main sources of physics background are \ttWj, \ttZj and \ttHj
processes. Smaller backgrounds,
for which all selected leptons are from $W$ or $Z$ boson decays or from leptonic $\tau$-lepton decays,
include diboson or triboson production, single-top-quark production in association with a
$Z$ boson (\tWZ, \tZq) and rare processes such as \ttt, \ttHH, \ttWH and \ttVV, where $V=W,Z$. These backgrounds, except for \ttWj
production, are evaluated using simulated events.
The \ttWj background is evaluated based on the information from the simulation and a data-driven technique described in
Section~\ref{sec:background}.
This is motivated because the \ttWj process is notoriously challenging to model accurately using only simulation~\cite{Frederix:2017wme}, and because
existing measurements of its cross section~\cite{TOPQ-2018-05,CMS-TOP-21-011,TOPQ-2016-11} are consistently above the theoretical predictions.
The background events originating from \ttjets and $tW$+jets production pass the selection if prompt leptons have a mis-assigned charge (\Qmis) or leptons are \fake. The \Qmis background is evaluated using a data-driven technique while the \fake background estimation is based on the template method
utilising inputs from both simulation and data as
described in Section~\ref{sec:background} and already used in Ref.~\cite{TOPQ-2018-05}.
 
The estimated yield for each source of background is given in Section~\ref{sec:results}.


\section{Data-driven background estimation}
\label{sec:background}

\subsection{Estimation of the \texorpdfstring{\ttW}{ttW} background}
\label{sec:background_ttW}
The theoretical modelling of the \ttW background at high jet multiplicity, corresponding to the phase space of this analysis, suffers from large uncertainties. To mitigate this effect, the normalisation of this background in jet multiplicity bins is determined using a data-driven approach, while other characteristics of the \ttW events are modelled by simulated events.
The uncertainties related to the modelling of the kinematic distributions
taken from simulation are discussed in Section~\ref{subsec:unc_irreducible}.
The estimate of the normalisation in each jet bin is based on a functional form that describes the evolution of
the number of \ttW events $N$ as a function of the jet multiplicity $j$,
$R(j)=N(j+1)/N(j)$~\cite{ELLIS1985435,BERENDS1989237,GIELE199014,Gerwick:2012hq}.
At high jet multiplicity, $R(j)$ follows the so-called 'staircase scaling' with $R(j)$ being a constant denoted $a_0$. This behaviour implies a fixed probability of additional jet radiation.
For lower jet multiplicities, Poisson scaling is expected~\cite{Gerwick:2012hq}
with $R(n) = a_1 / (1+n)$, where $a_1$ is a constant and $n$ is the number of additional jets to the hard process. It was derived in the Abelian limit of QCD by considering gluon radiation off hard quark legs and validated using experimental data~\cite{Gerwick:2012hq}.
The transition point between these scaling behaviours depends on the jet kinematic selection.
Since \ttW production in the phase-space region of this analysis is dominated by events that include at least four jets in the matrix element at tree level, the parameterisation of the \ttW events starts at the fourth jet ($n=j-4$).
Independent normalisation factors (NF) for the expected number of $t\bar{t}W^{+}$ and $t\bar{t}W^{-}$ events are introduced to take into account different production rates of these processes in \pp collisions.
The normalisation factor for \ttW events with $j>4$ is then given by
\begin{equation}
\mathrm{NF}_{t\bar{t}W(j)} = \mathrm{NF}_{ t\bar{t}W^+ (\mathrm{4jet})}\times \Pi_{j'=4}^{j'=j-1} \left[a_0+\frac{a_1}{1+(j'-4)}\right] + \mathrm{NF}_{t\bar{t}W^- (\mathrm{4jet})}\times   \Pi_{j'=4}^{j'=j-1} \left[a_0+\frac{a_1}{1+(j'-4)} \right].\label{equation:ttW}
\end{equation}
Dedicated control regions (CR) labelled as \CRttWp, \CRttWm, \CROnebp and \CROnebm are defined to determine the $a_0$ and $a_1$ scaling parameters and the overall number of $t\bar{t}W^+$ and $t\bar{t}W^-$ events with exactly 4 jets, $\mathrm{NF}_{ t\bar{t}W^+ (\mathrm{4jet})}$ and $\mathrm{NF}_{ t\bar{t}W^- (\mathrm{4jet})}$, respectively. The definition of these CRs is summarised in Table~\ref{tab:CRdef}.
The \CRttWp and \CRttWm regions are primarily used
to adjust the \ttW normalisation in a phase space with at least two $b$-tagged jets as in the SR, while
the \CROnebp and \CROnebm regions, which include events with $N_j \ge 4$, are mainly used to
determine the \ttW jet multiplicity spectrum.
 
The \ttW CRs, along with the CRs defined in Section~\ref{sec:background_fake}, are included in the likelihood fit to data, together with the SR (see Section~\ref{sec:signal}).
\Cref{fig:ttWCR} shows the jet-multiplicity distributions in each of the four \ttW CRs. After the fit, the predictions provide a good description of the data.
 
As mentioned above, unlike the signal and other background processes, \ttW production has a pronounced charge asymmetry in \pp collisions: $t\bar{t}W^{+}$ events are produced approximately twice as abundantly as $t\bar{t}W^{-}$ events. This property is used to validate the data-driven estimation of this background by examining the difference between the number of events with all tight leptons positively charged (positive events) and the number of events with all tight leptons negatively charged (negative events). The charge-symmetric \tttt\ signal and most non-\ttW backgrounds are expected to cancel in this difference, providing a validation of the \ttW background. The distribution of the number of jets obtained from this procedure for events entering the four \ttW CRs
and the SR defined in Table~\ref{tab:CRdef} is shown in Figure~\ref{fig:ttWVR}.
A good description of the jet multiplicity distribution in data confirms that it can be predicted by two scaling parameters, as given in Equation~\ref{equation:ttW}.
\begin{figure}[htbp]
\centering
\subfloat[]{\includegraphics[width=0.43\linewidth]{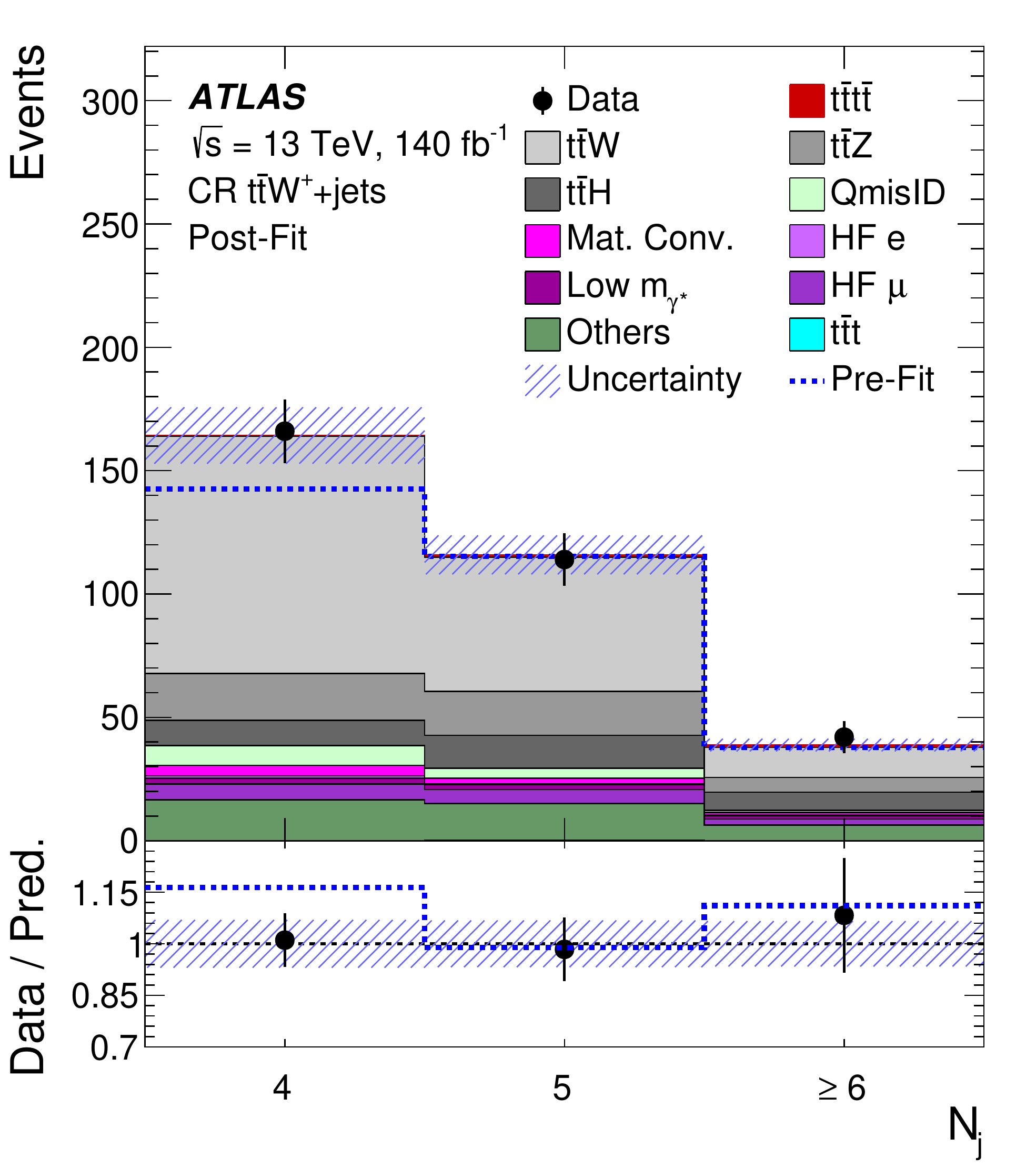}}
\subfloat[]{\includegraphics[width=0.43\linewidth]{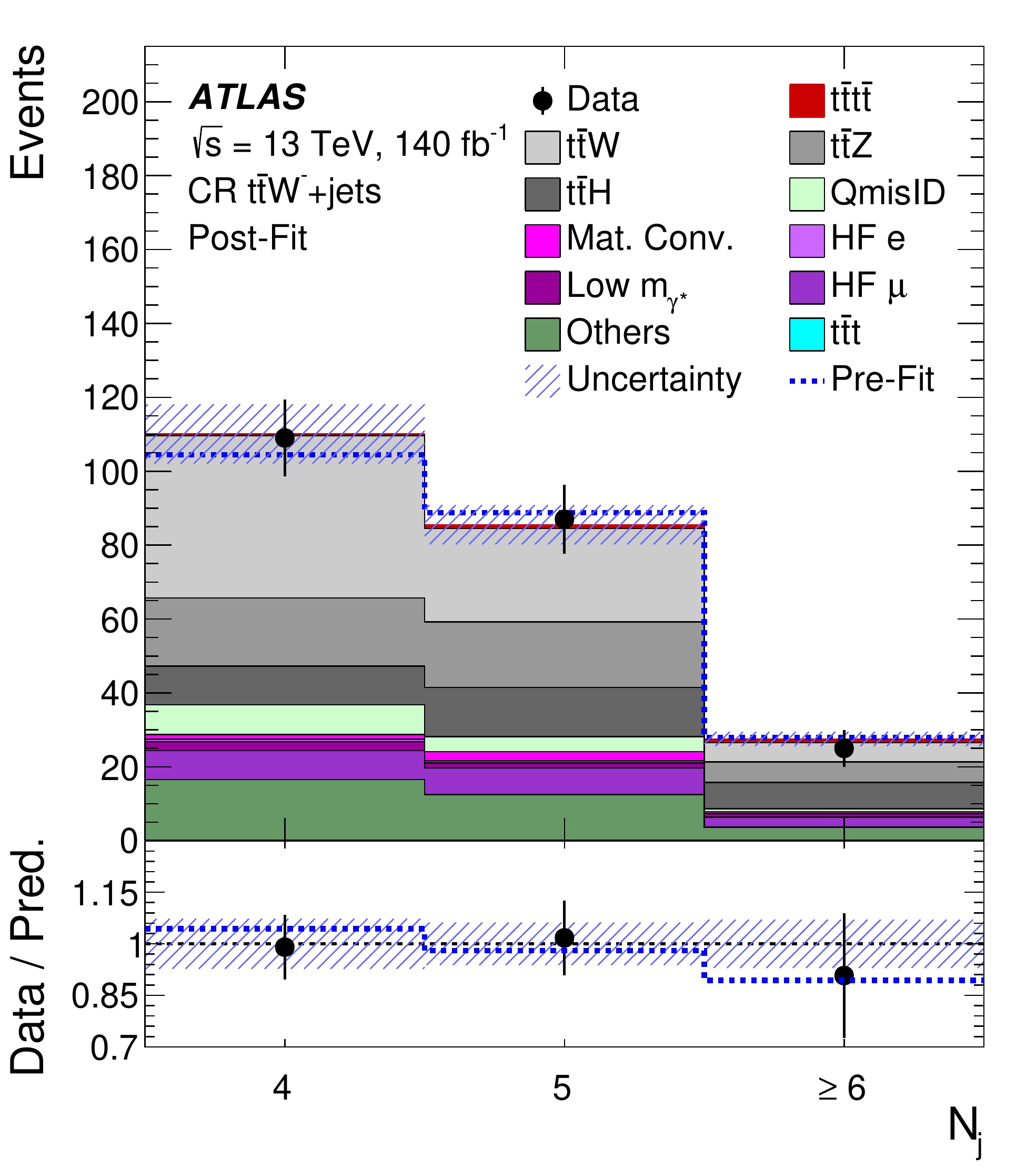}}
\\
\subfloat[]{\includegraphics[width=0.43\linewidth]{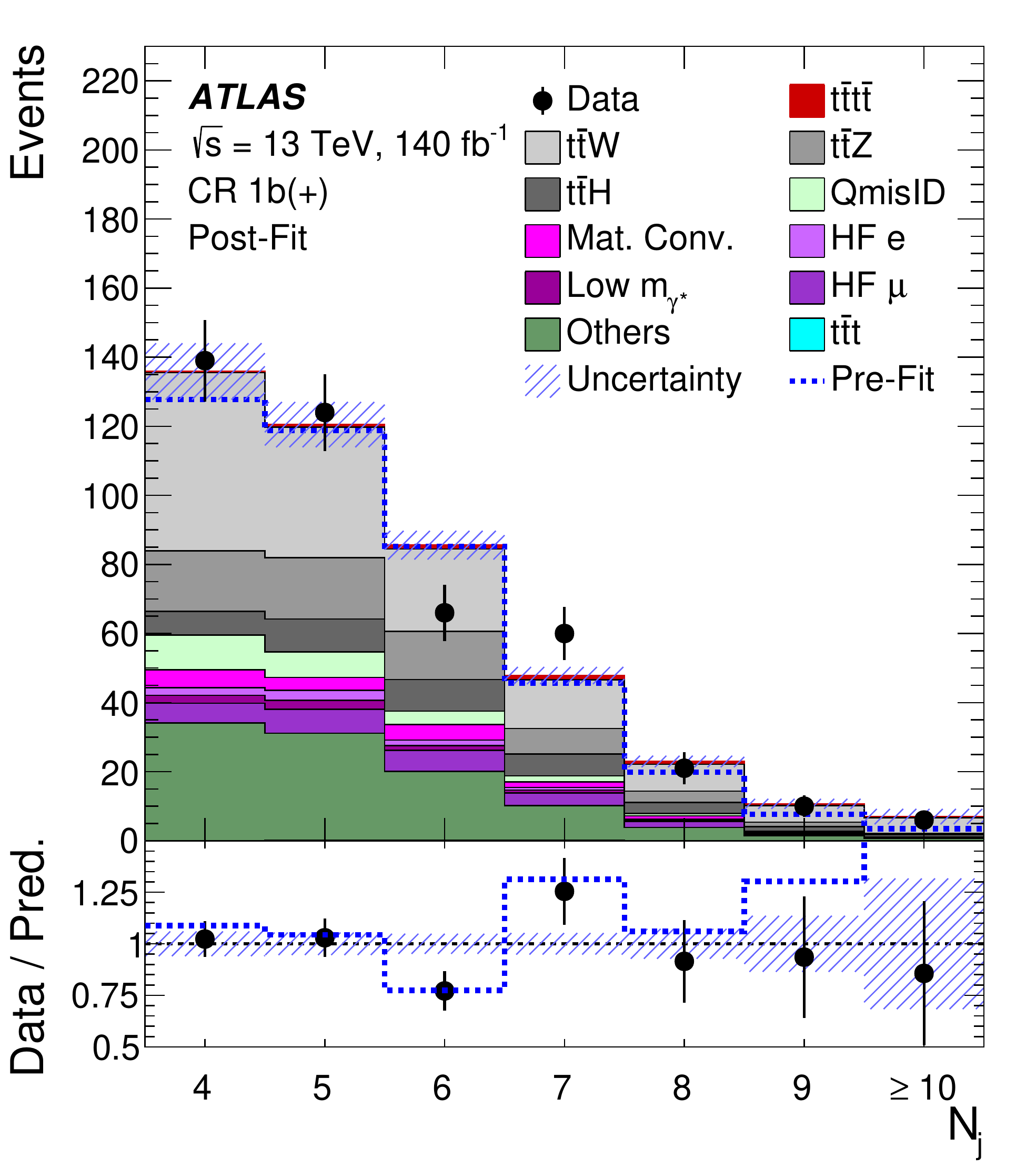}}
\subfloat[]{\includegraphics[width=0.43\linewidth]{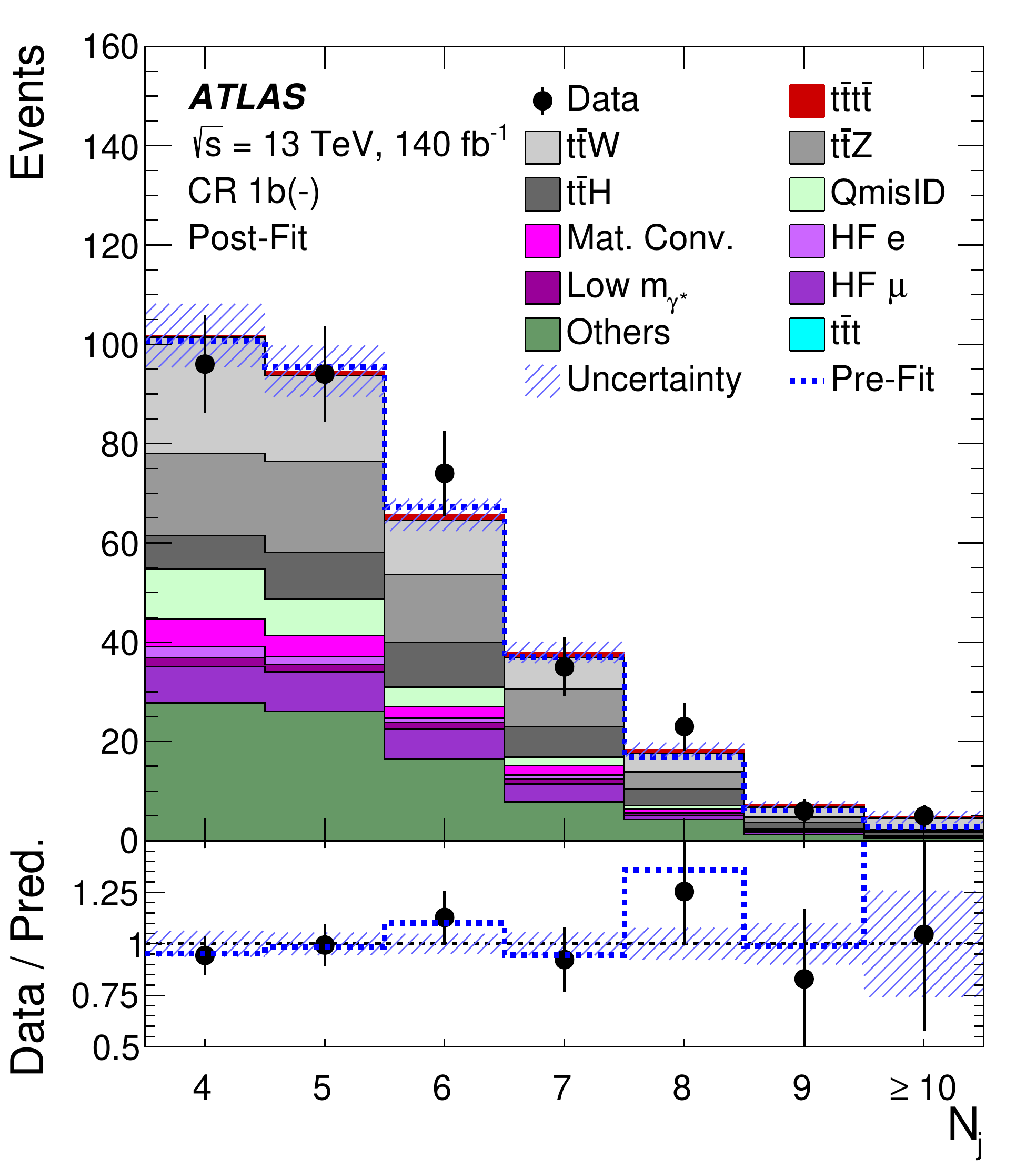}}
\caption{\label{fig:ttWCR} Post-fit distributions for the number of jets ($N_{j}$) in each of the four \ttW CRs:  (a) \CRttWp, (b) \CRttWm, (c) \CROnebp and (d) \CROnebm. The ratio of the data to the total post-fit prediction is shown in the lower panel.
The dashed blue lines show the pre-fit prediction in the upper panel and the ratio of the data to the total pre-fit prediction in the lower panel. The shaded band represents the total post-fit uncertainty in the prediction.
}
\end{figure}
 
\begin{figure}
\centering
\includegraphics[width=0.5\linewidth]{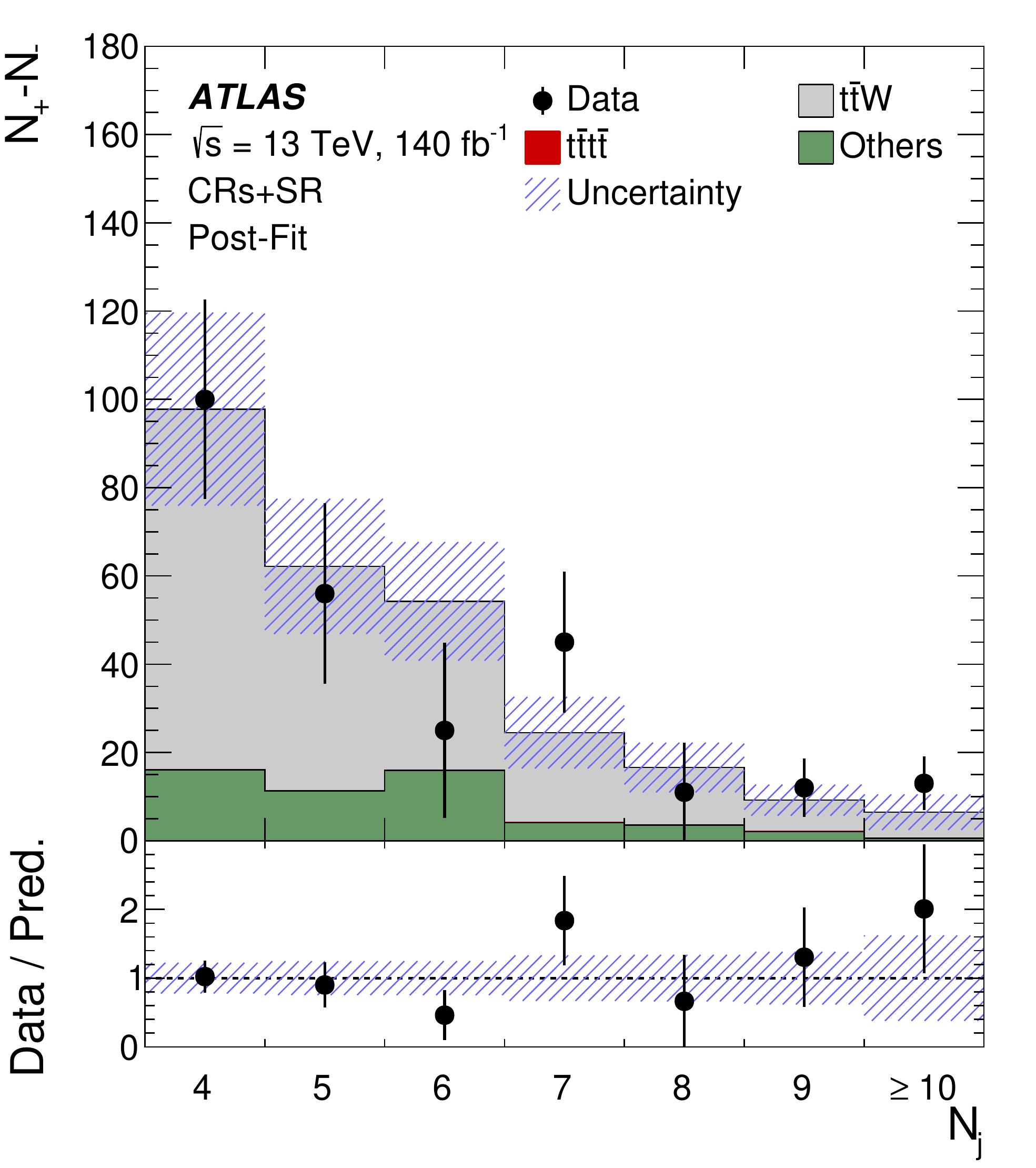}
\caption{\label{fig:ttWVR}
Post-fit distribution for the difference between the number of positive events and the number of negative events ($N_+-N_-$) as a function of the number of jets (\njets) in the sum of four \ttW CRs and the SR. The uncertainties on the normalisation factors and on the \ttW modelling parameters are represented by the shaded band. The ratio of the data to the total post-fit prediction is shown in the
lower panel.
}
\end{figure}

\begin{table}[!htb]
\centering
\caption{Definition of the signal and control regions. The control regions labeled as \CRttWp, \CRttWm, \CROnebp and \CROnebm are defined to  determine the background modelling parameters of $t\bar{t}W^+$ and $t\bar{t}W^-$ events. The control regions labeled as \CRHFe, \CRHFm, \CRCO and \CRGst are defined to determine the normalisation of fake/non-prompt lepton background events. \njets ($N_b$) indicates the jet ($b$-tagged jet) multiplicity in the event. \Ht is defined as the scalar sum of the transverse momenta of the isolated leptons and jets.
$\ell_{1}$, $\ell_{2}$ and $\ell_{3}$ refer to the highest \pt, second-highest \pt and third-highest \pt leptons, respectively.
$\eta(\text{e})$ refers to the electron pseudorapidity. Total charge is the sum of charges for all leptons. GNN denotes the Graph Neural Network discriminant trained to separate \tttt signal from the background described in Section~\ref{sec:signal}.
}
\footnotesize{
\begin{tabular}{ c | c | c | c | c | c }
\hline
\multirow{2}{*}{Region} & \multirow{2}{*}{Channel} & \multirow{2}{*}{\njets} & \multirow{2}{*}{$N_{b}$} & Other & Fitted \\
& & & & selection & variable\\
\hline\hline
\multirow{2}{*}{\CRGst} & \multirow{2}{*}{SS, ee or e$\mu$} & \multirow{2}{*}{$4 \le \njets< 6$} & \multirow{2}{*}{$\ge 1$} &  $\ell_{1}\; \mathrm{or}\; \ell_{2}$ is from virtual photon ($\gamma$*) decay & \multirow{2}{*}{event yield}  \\
& & & & $\ell_{1} \; \mathrm{and}\; \ell_{2}$ are not from photon conversion & \\
\hline
\CRCO & SS, ee or e$\mu$ & $4 \le \njets< 6$ & $\ge 1$ &  $\ell_{1}\; \mathrm{or}\; \ell_{2}$ is from photon conversion & event yield \\
\hline
\multirow{4}{*}{\CRHFm} & \multirow{4}{*}{e$\mu\mu$ or $\mu\mu\mu$} & \multirow{4}{*}{$\ge 1$} & \multirow{4}{*}{$= 1$} & $100 < \Ht < 300$ \GeV\ & \multirow{4}{*}{\pTthirdleadlep}\\
& &  & & $\met > 50$ \GeV\ & \\
& &  & &  total charge $= \pm$1 & \\
\hline
\multirow{4}{*}{\CRHFe} & \multirow{4}{*}{eee or ee$\mu$} &\multirow{4}{*}{$\ge 1$} & \multirow{4}{*}{$= 1$} & $100 < \Ht < $ 275 \GeV\  & \multirow{4}{*}{\pTthirdleadlep}\\
& &  & & $\met > 35$ \GeV\ & \\
& &  & & total charge $= \pm$1 & \\
\hline
\multirow{5}{*}{\CRttWp} & \multirow{5}{*}{SS, e$\mu$ or $\mu\mu$} & \multirow{5}{*}{$\ge 4$}  & \multirow{5}{*}{$\ge 2$} &  $ |\eta(e)| < 1.5$ & \multirow{5}{*}{$\njets$} \\
& & & & when $N_{b}=2$: $ \Ht< 500$ \GeV\ or $ N_{{j}} < 6$ & \\
& & & &  when $N_{b}\ge 3$: $ \Ht < 500$ \GeV\ & \\
& & & &  total charge $>0$   & \\
\hline
\multirow{5}{*}{\CRttWm} & \multirow{5}{*}{SS, e$\mu$ or $\mu\mu$} & \multirow{5}{*}{$\ge 4$}  & \multirow{5}{*}{$\ge 2$} & $ |\eta(e)| < 1.5$ & \multirow{5}{*}{$\njets$} \\
& & & & when $N_{b}=2$: $ \Ht< 500$ \GeV\ or $ N_{{j}} < 6$ & \\
& & & &  when $N_{b}\ge 3$: $ \Ht < 500$ \GeV\  & \\
& & & &  total charge $<0$   & \\
\hline
\multirow{3}{*}{\CROnebp} & \multirow{3}{*}{2LSS+3L} & \multirow{3}{*}{$\ge 4$} & \multirow{3}{*}{$ = 1$} & $\ell_{1}$ and $\ell_{2}$ are not from photon conversion   & \multirow{3}{*}{$\njets$}\\
& & & & $\Ht >500$ \GeV\  & \\
& & & & total charge $>0$  & \\
\hline
\multirow{3}{*}{\CROnebm} & \multirow{3}{*}{2LSS+3L} & \multirow{3}{*}{$\ge 4$} & \multirow{3}{*}{$ = 1$} & $\ell_{1}$ and $\ell_{2}$ are not from photon conversion  & \multirow{3}{*}{$\njets$}\\
& & & & $\Ht >500$ \GeV\  & \\
& & & & total charge $<0$  & \\
\hline
\hline\hline
SR & 2LSS+3L & $\ge 6$ & $ \ge 2$ & $\Ht >500$ \GeV\ & GNN score \\
\hline
\end{tabular}
}
\label{tab:CRdef}
\end{table}


\begin{figure}[htbp]
\centering
\subfloat[]{\includegraphics[width=0.43\linewidth]{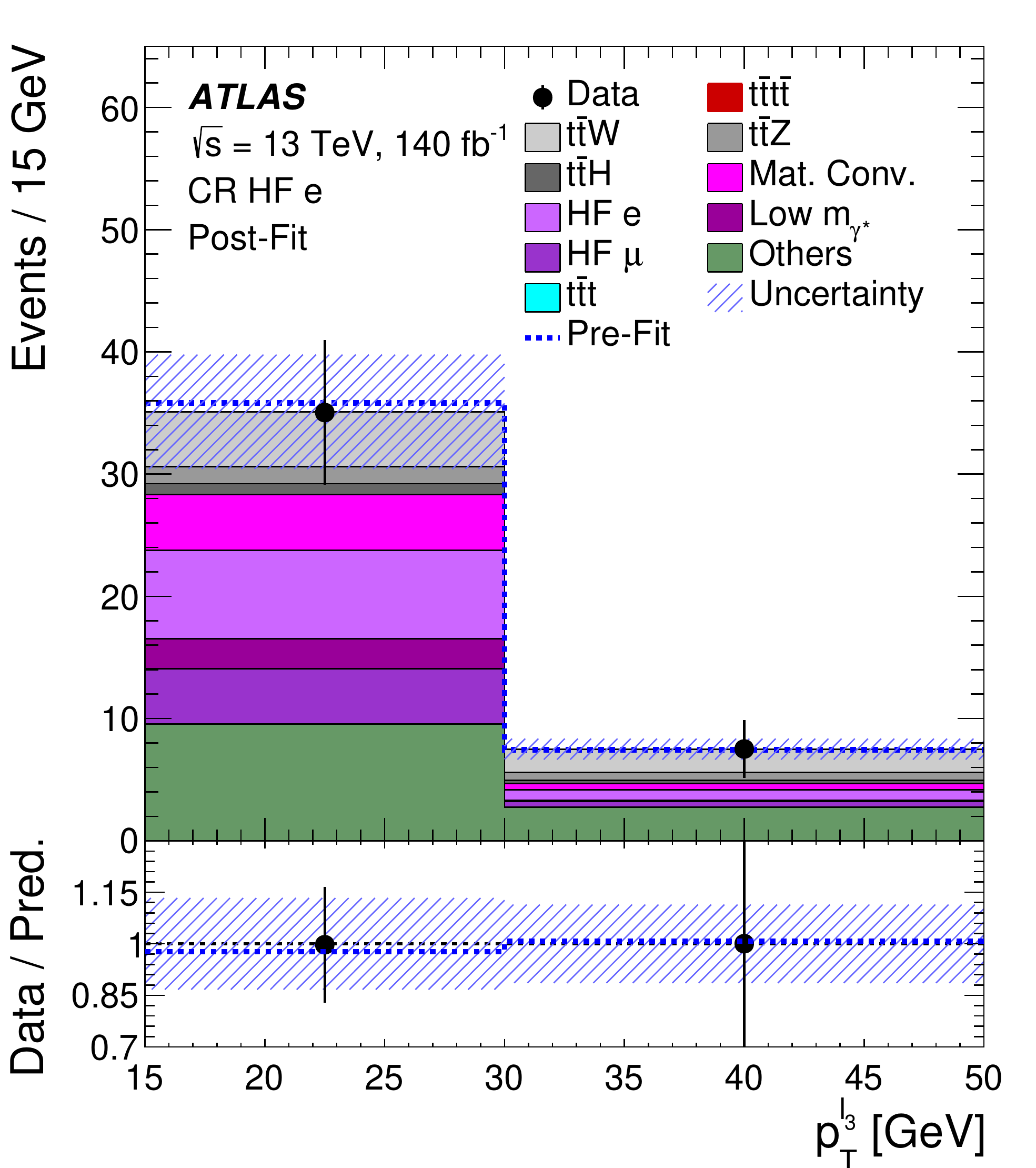}}
\subfloat[]{\includegraphics[width=0.43\linewidth]{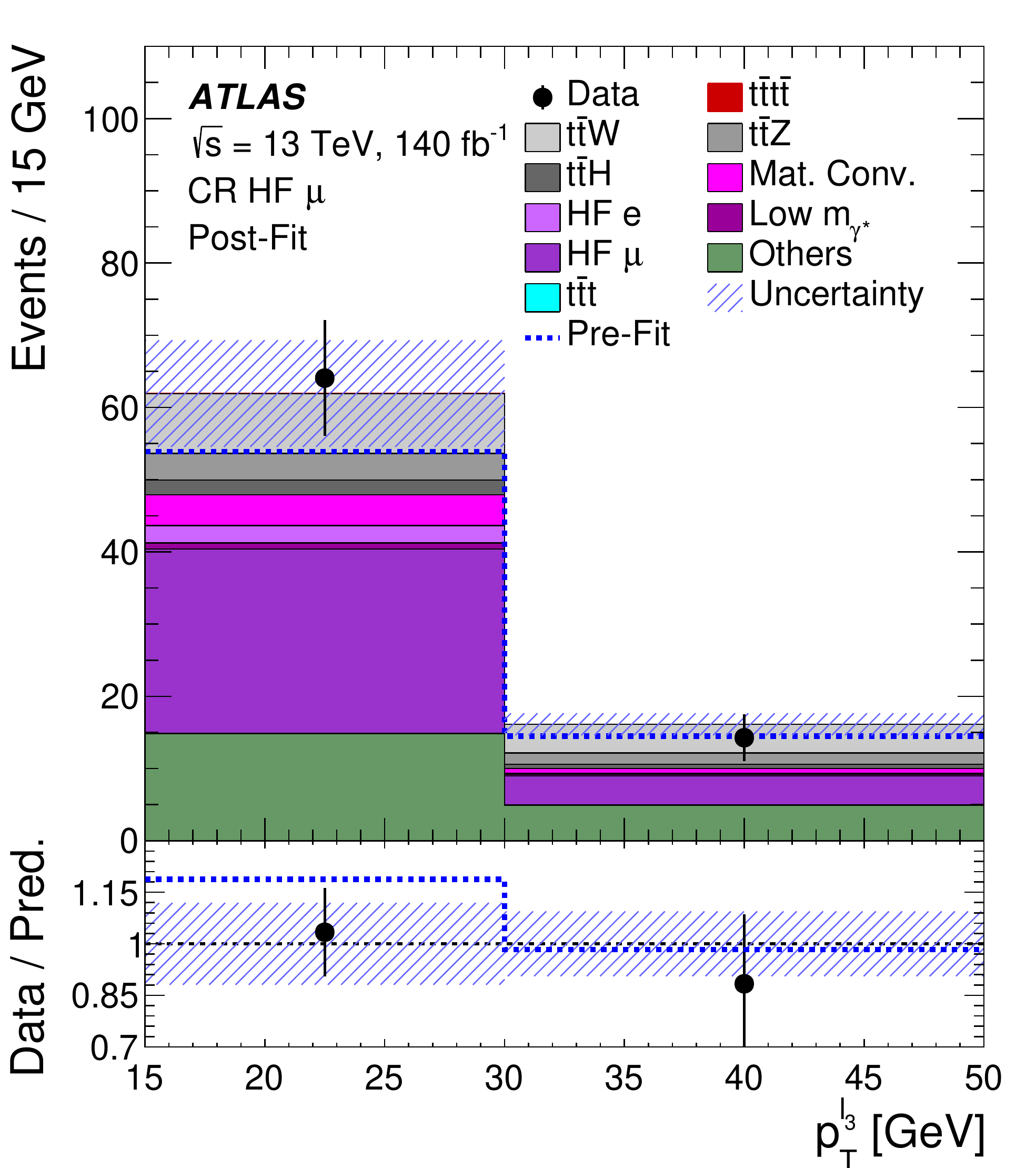}}\\
\subfloat[]{\includegraphics[width=0.43\linewidth]{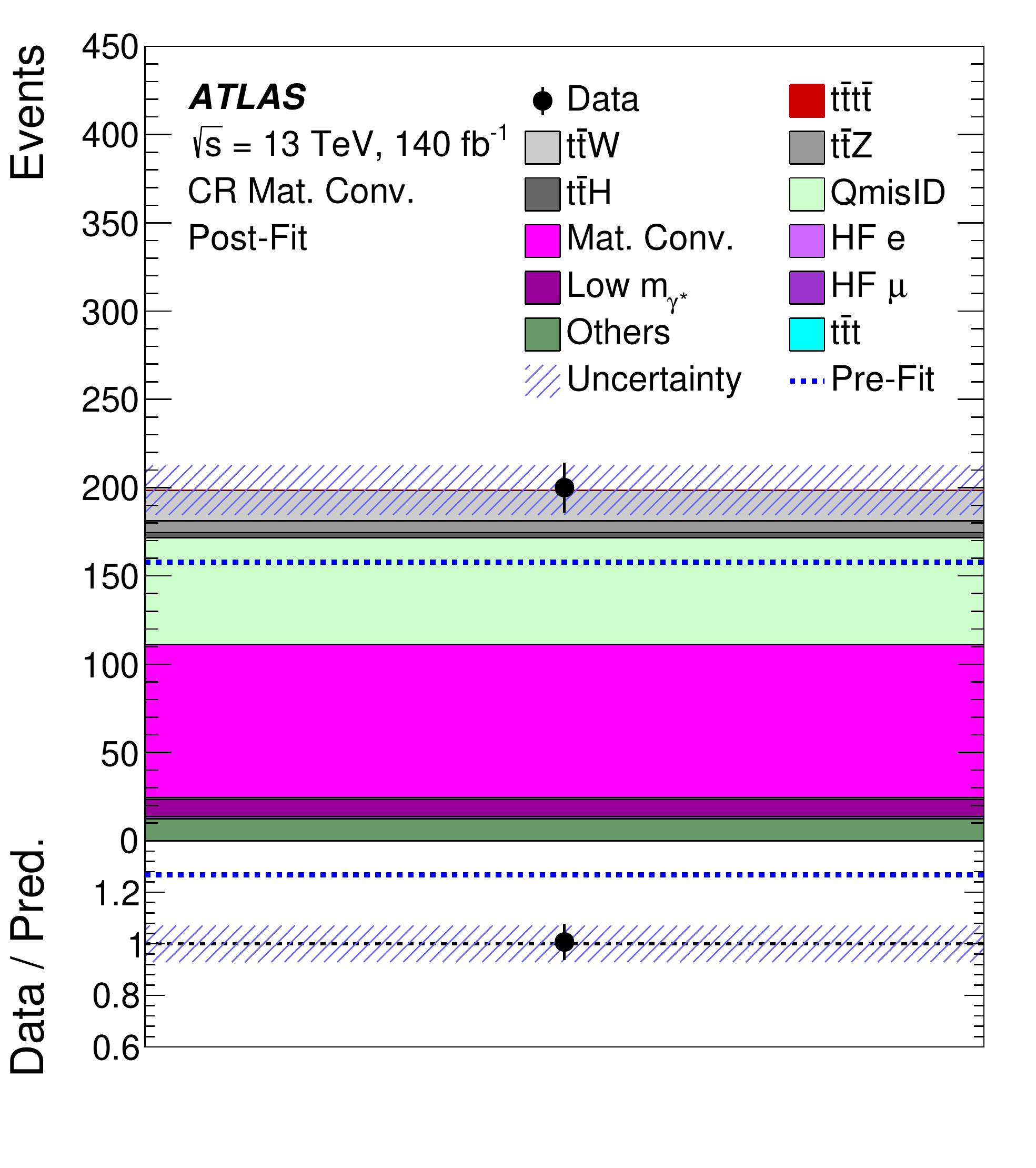}}
\subfloat[]{\includegraphics[width=0.43\linewidth]{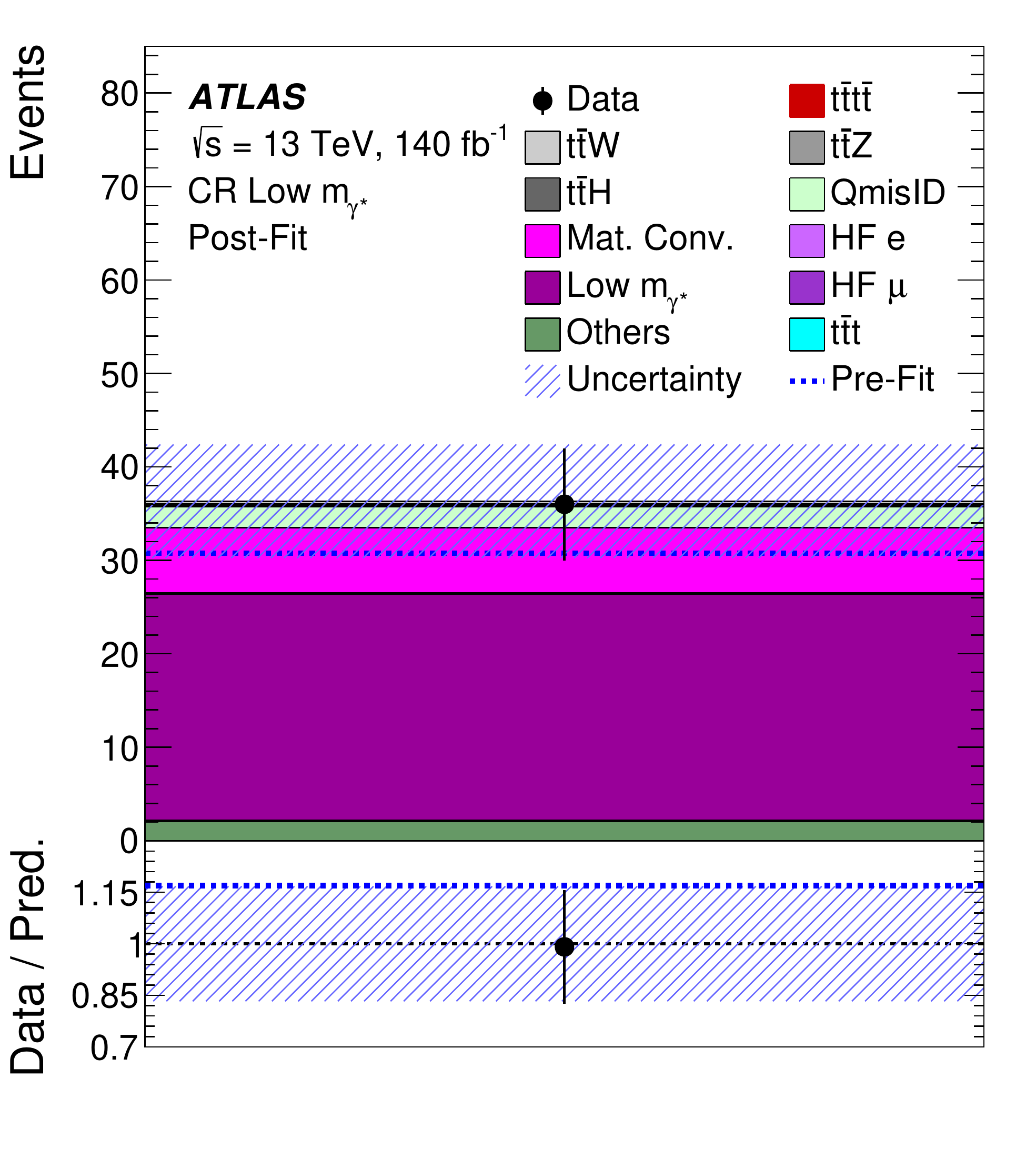}}
\caption{\label{fig:fit:real:PostFit} Post-fit distributions for the fitted variables in the CRs for the fake/non-prompt lepton background. For (a) \CRHFe and (b) \CRHFm, the third-highest lepton \pt is fitted, while for (c) \CRCO and (d) \CRGst, the number of events is used. The ratio of the data to the total post-fit prediction is shown in the lower panel. The dashed blue lines show the pre-fit prediction in the upper panel and the ratio of the data to the total pre-fit prediction in the lower panel. The shaded band represents the total post-fit uncertainty in the prediction. In (a) and (b), the last bin contains the overflow.}
\end{figure}
 
\subsection{Fake/non-prompt lepton background}
\label{sec:background_fake}
A template method is used to estimate the fake/non-prompt background.
This method relies on simulation samples to model the kinematic distributions of the background processes arising from fake and non-prompt leptons and on CRs to determine the normalisation of these components.
These CRs are included in the fit, together with the \ttW CRs and the SR, and the normalisation factors are determined simultaneously with the \tttt signal.
The uncertainties related to the modelling of the kinematic distributions predicted in the simulation are discussed in Section~\ref{subsec:unc_reducible}.
The method is similar to that used in the previous analysis~\cite{TOPQ-2018-05}.
 
The following main contributions of the fake/non-prompt background are distinguished:
\begin{itemize}
\item Events with a single reconstructed non-prompt electron (muon) from heavy-flavour
decay; this contribution is referred to as  \HFe (\HFm).
\item Events with a single reconstructed non-prompt electron originating from photon conversions in the detector material (\CO).
\item Events with a virtual photon ($\gamma^*$) leading to an $e^{+}e^{-}$ pair where only one of the electrons is reconstructed (\Gst).
\end{itemize}
Minor contributions to the fake/non-prompt background arising from events with a lepton originating from light-meson decay or with a jet mis-identified as a lepton are determined using simulated events.
 
As summarised in Table~\ref{tab:CRdef}, the CRs labelled as \CRHFe and \CRHFm are defined to determine the normalisation of the \HFe and \HFm backgrounds, respectively. Similarly, the CRs labelled as \CRCO and \CRGst are defined to determine the normalisation of \CO and \Gst backgrounds, respectively.
Each region is designed to have a dominant background component.
 
Figure~\ref{fig:fit:real:PostFit} shows the distributions and event yields used in the template method: the third-highest lepton \pt and the number of events.
A good description of the data distributions by the fitted predictions is observed in all CRs.
 
\subsection{Charge mis-assignment background}
The QmisID background affects only the $ee$ and $e\mu$ channels of the 2LSS region, and is estimated using the same method as in the previous analysis~\cite{TOPQ-2018-05}.
The probability for an electron to have its charge incorrectly assigned is estimated using a data sample of $Z\rightarrow ee$ events requiring the invariant mass of the electron pair to be within 10~\gev\ of the $Z$ boson mass and without any requirement on the charge of the two electrons. The charge mis-assignment rate extracted from the fraction of events with same-charge electrons within this sample,
is parameterised as a function of the electron transverse momentum and pseudorapidity.
The rate varies from 0.004\% to 4\% depending on the electron \pt and $|\eta|$.
The expected number of events arising from the QmisID background is determined by applying the measured charge mis-assignment rate to data events satisfying the requirements of the kinematic selection of the 2LSS channel, except that the two leptons are required to be of opposite charge.


\section{Signal extraction and cross section measurement}
\label{sec:signal}
 
The background composition of the SR is largely dominated by the production of top-quark pairs in
association with bosons. A multivariate discriminant built with a Graph Neural Network (GNN)~\cite{Battgalia_2018} is used to separate the \tttt\ signal from the background using the \textsc{graph\_nets} library from TensorFlow~\cite{Abadi:2016kic}.
From each event, a fully connected graph is constructed with `nodes', corresponding to reconstructed jets, electrons, muons, and the missing transverse momentum of the event. The features of each node include the four momentum of the object, assumed to be massless,\footnote{The energy of each object is computed from the four momentum
assuming that its mass is zero. For jets, this avoids any dependence on the jet mass calibration.} the jet PCBT score, the lepton charge, and an integer labelling the type of object represented by the node. The `edges' between nodes carry three features
with information about the angular separation between the objects they connect. Additionally, the jet multiplicity is treated as a `global' feature. The nodes, edges, global feature and the GNN hyperparameters are optimised to maximise the integral under the receiver operating characteristic curve of the GNN event classifier.

The GNN training is performed for events passing the SR requirements. The LO \tttt simulated signal sample is used in the training. The MC simulated samples,
corresponding to all background components, represent the background in the training.
The GNN discriminant is chosen as the observable of the analysis to extract the \tttt signal.
Figure~\ref{fig:SR_GNN} shows the distribution of the GNN score in the signal region.

Following the strategy of the previous publication~\cite{TOPQ-2018-05}, a BDT discriminant is also trained as a cross-check in the SR by combining several input observables, of which the sum of the four highest PCBT scores among all jets in the event has the highest discriminating power.
The separation power of the \tttt  signal from the background is slightly better for the GNN, leading to a 10\% higher expected significance for this method.
 
The \tttt production cross-section, the normalisation factors of the backgrounds, and the \ttW modelling parameters defined in Section~\ref{sec:background_ttW} are determined via a binned likelihood fit to the discriminant  score distribution in the SR and to the event yields or to the discriminating variable distributions in the eight CRs listed in Table~\ref{tab:CRdef}. The maximum-likelihood fit is performed with the RooFit package~\cite{RooFit} based on a likelihood function built on the Poisson probability that the observed data are compatible with the model prediction. The value of each nuisance parameter, describing the systematic uncertainties for both signal and background processes (see Section~\ref{sec:systematics}), is constrained by a Gaussian penalty term present in the likelihood function, while all normalisation factors and \ttW modelling parameters, described in Section~\ref{sec:background}, are unconstrained.


\section{Systematic uncertainties}
\label{sec:systematics}
 
Systematic uncertainties arise from the modelling of the signal processes,
from theoretical or data-driven predictions of the background processes,
as well as from experimental effects.
These uncertainties affect both the shape and the normalisation of the estimations.
In addition, they can lead to event migration between different regions.
The different sources of systematic uncertainty are described below and their impact on the \tttt cross section measurement after the fit is shown in Table~\ref{tab:syst}.
 
\subsection{Experimental uncertainties}
 
The dominant experimental uncertainties arise from the measurement of the $b$-tagging efficiencies and mis-tagging rates~\cite{FTAG-2018-01,FTAG-2020-08,ATLAS:2023lwk}.
Corrections applied to simulated samples to match the performance in data and the associated uncertainties
are determined for each jet flavour separately in five PCBT score bins and in bins of jet \pt. The resulting uncertainties are decomposed into 40 independent components (referred to as eigenvector components)
for $b$-jet tagging, 20 for $c$-jet and 20 for light-jet mis-tagging.
Due to a lack of data for $b$-/$c$-/light jets calibration with $\pt>400/300/250$~\GeV,
the corrections for these jets are extrapolated from the highest \pt\ bin below these thresholds;
uncertainties in this extrapolation are determined from the simulation, independently for $b$-/$c$-/light jets.
 
Uncertainties in the calibration of the jet energy scale and resolution~\cite{JETM-2018-05,PERF-2011-03,PERF-2011-04} play a subleading role among the experimental uncertainties.
An additional minor uncertainty arises from the correction applied to simulated events associated with the jet-vertex-tagger selection~\cite{PERF-2014-03}.
 
Other experimental uncertainties have minor impacts on the measurements.
The uncertainty in the combined 2015--2018 integrated luminosity is 0.83\% \cite{DAPR-2021-01},
obtained using the LUCID-2 detector~\cite{LUCID2},
complemented by measurements using the inner detector and calorimeters.
An uncertainty in the corrections on the pile-up profile in the simulated samples is also considered.
Uncertainties in electrons and muons arise from the calibration of the simulated samples to data.
The calibrations correct for the efficiencies of the trigger, reconstruction, identification and isolation requirements,
as well as the energy scale and resolution~\cite{EGAM-2018-01,CERN-EP-2020-199}.
For electrons, an additional uncertainty associated with the efficiency of the electric-charge identification is considered.
Finally, an uncertainty in the measurement of \met\ is assigned due to a possible mis-calibration of its soft-track component~\cite{PERF-2016-07}.
 
\subsection{Signal modelling uncertainties}
 
Uncertainties in the modelling of SM \tttt\ production have the dominant impact on the measurements.
They affect both the signal acceptance and the shape of the GNN discriminant.
The two leading uncertainties are determined by comparing the nominal prediction with alternative samples generated with \SHERPA\ and with \mgamcc+\herwigseven.
These uncertainties mainly cover two effects: different matching schemes between the matrix element and parton shower generators as well as different parton shower, hadronisation and fragmentation models.
The uncertainty due to missing higher-order QCD corrections in simulated events is estimated
by varying the renormalisation and factorisation scales in the matrix element individually or simultaneously up or down by a factor of two,
and taking the maximum up and down variations.
The PDF uncertainty is 1\%.
It is calculated as the RMS of the predictions from the 100 replicas of the \texttt{NNPDF30\_nlo\_as\_0118} PDF set following the PDF4LHC prescription~\cite{Butterworth:2015oua}.
The effect of PDF variation on the shape of the fitted distributions was found to be negligible.
 
\subsection{Uncertainties in irreducible backgrounds}
\label{subsec:unc_irreducible}
The dominant uncertainty in the background predictions arises from the modelling of \ttW, \ttH\ and \ttZ\ events.
Uncertainties due to generator choices are determined by comparing the
nominal predictions with those obtained with the alternative matrix element and/or parton shower generators specified in Table~\ref{tab:mc_config}.
Given that the \njets\ distribution in \ttW\ production is estimated using a data-driven method,
the alternative MC sample simulated using \mgamc\ is  weighted to have the same \njets\ distribution as the nominal prediction.
This avoids double counting of the systematic effects in \njets\ modelling.
Minor uncertainties due to missing higher-order QCD corrections and PDF variations for each of the three processes are studied and determined using the same method as for \tttt\ production.
 
The uncertainty in the predicted cross-sections of the background processes have minor impact on the measurements.
The uncertainty in the \ttt\ production cross section is set to 35\%.
This includes the effect of varying the renormalisation and factorisation
scales up and down by a factor of two in the NLO QCD prediction,
and the uncertainties from the PDF choice, EW LO contribution and the acceptance difference between MC samples generated using the five-flavour and the four-flavour schemes.
 
Uncertainties of 12\% and 10\% are included for the \ttZ\ and \ttH\ production cross sections, respectively~\cite{deFlorian:2016spz}.
An uncertainty of 30\% is applied to the sum of \tZq\ and \tWZ\
processes~\cite{TOPQ-2016-14,Faham:2021vde,BessidskaiaBylund:2016bqj}. 
The uncertainty in $VV$ production is taken as the discrepancies between the measured differential cross section as a function of the jet multiplicity and the prediction from \SHERPA2.2.2~\cite{STDM-2018-03}.
For $VV$ events with $\leq3/4/\geq5$ jets, an uncertainty of 20\%/50\%/60\% is applied.
The three components are considered as uncorrelated.
An uncertainty of 50\% is considered for \ttWW\ production based on the NLO prediction~\cite{Alwall:2014hca}.
For all other minor background processes with negligible impact including the production of $VH$, $VVV$, \ttZZ, \ttWZ, \ttHH\ and \ttWH,
a single uncertainty of 50\% is considered.
This value is based on previous analyses in a similar final state~\cite{TOPQ-2018-05,EXOT-2016-16},
and is large enough to cover the different predictions listed in Ref.~\cite{Alwall:2014hca}.
 
For all background contributions estimated based on simulations,
additional uncertainties are considered for events produced with associated $b$-jets.
This is motivated by the expected high $b$-jet multiplicities in signal-like events
and the general difficulty in modelling additional heavy-flavour jets in simulations.
For each background process, an uncorrelated uncertainty of 50\% is assigned to events
with exactly three `true' $b$-jets
\footnote{
`True' $b$-jets are identified in simulated events as jets that are ghost-associated
with $b$-hadrons~\cite{Cacciari:2008gp}.
}
and to events with at least four true $b$-jets.
Since \ttt events have three $b$-jets arising from top quark decays, an uncertainty of 50\% is only assigned to \ttt events if they have at least four true $b$-jets.
These estimates are based on the measurement of
\ttbar production with additional heavy-flavour jets~\cite{TOPQ-2017-12}
and on comparisons between data and prediction in $t \bar{t} \gamma$
events with three and four $b$-tagged jets.
 
\subsection{Uncertainties in reducible backgrounds}
\label{subsec:unc_reducible}
Uncertainties in the estimate of the fake/non-prompt and QmisID backgrounds have minor impact.
These uncertainties are derived and treated in the same way as in Ref.~\cite{TOPQ-2018-05}.
The resulting uncertainties for the QmisID background range from 5-30\% depending on the \pt\ and $\eta$ of the lepton.
For leptons from heavy-flavour hadron decay, the uncertainties range from 20-100\%.
For material conversion and virtual photon conversion, the uncertainties are 30\% and 21\%, respectively.
Conservative uncertainties are applied to other smaller contributions,
including an uncertainty of 100\% for events with \fake\ leptons from light jets,
and a 30\% uncertainty for all other sources of \fake\ background.
These values are taken from previous analyses in a similar final state~\cite{TOPQ-2018-05,EXOT-2016-16}.
 
Given that \ttbar\ events are the main source of fake/non-prompt and QmisID backgrounds,
additional uncertainties are considered for \ttbar\ events produced in association with additional $b$-jets.
An uncorrelated uncertainty of 50\% is assigned to events with 3 $b$-jets and to events with at least four $b$-jets.


\section{Result for the \texorpdfstring{\tttt}{tttt} cross section measurement}
\label{sec:results}
A maximum-likelihood fit is performed to the GNN score distribution in the SR and the different distributions used in the CRs (see Table~\ref{tab:CRdef}). Figure~\ref{fig:SR_GNN} shows the distribution of the GNN score in the SR before and after
performing the fit. Good agreement is observed between data and the prediction after the fit.
The best-fit value of the signal strength $\mu$, defined as the ratio of the measured \tttt cross section to the
SM expectation of $\stttt = 12.0 \pm 2.4 ~\text{fb}$ from Ref.~\cite{Frederix:2017wme},
is found to be:
\begin{equation}\nonumber
\mu=1.9 \pm 0.4 \mathrm{(stat)}\,^{+0.7}_{-0.4}\mathrm{(syst)}=1.9\,^{+0.8}_{-0.5}.
\end{equation}
The corresponding measured \tttt production cross section is:
\begin{equation}\nonumber
\sigma_{\tttt}=22.5^{+4.7}_{-4.3} \mathrm{(stat)}\,^{+4.6}_{-3.4}\mathrm{(syst)}\,\text{fb}=22.5\,^{+6.6}_{-5.5}\,\text{fb}.
\end{equation}
The systematic uncertainty, is determined by subtracting the statistical uncertainty in quadrature
from the total uncertainty.
The statistical uncertainty is obtained from a fit where all nuisance parameters are fixed to their post-fit values. The systematic uncertainty on the signal strength includes the theoretical uncertainty
on the SM \tttt cross section.
The measured \tttt production cross section is consistent within 1.8 standard deviations with
the SM prediction of Ref.~\cite{Frederix:2017wme},
and within 1.7 standard deviations with the resummed $\stttt$ calculation of Ref.~\cite{vanBeekveld:2022hty}.
 
The probability for the background-only hypothesis to result in a signal-like excess at least as large as seen in data is derived using the profile-likelihood ratio following the procedure described in Ref.~\cite{Cowan:2010js}.
From this, the significance of the observed signal is found to be 6.1 standard deviations, while 4.3 standard deviations are expected using the SM cross section of $\stttt = 12.0 \pm 2.4 ~\text{fb}$ from Ref.~\cite{Frederix:2017wme}.
Using the SM cross section of $13.4^{+1.0}_{-1.8}$~fb from Ref.~\cite{vanBeekveld:2022hty}, the expected significance would be 4.7 standard deviations.
The goodness-of-fit evaluated using a saturated model~\cite{Baker:1983tu,Saturated} yields a probability of 76\%.
Compared to the previous result of Ref.~\cite{TOPQ-2018-05}, the gain in expected sensitivity comes from the updated lepton and jet selection and uncertainties, from the use of the GNN discriminant and from the improved treatment of the \ttt background. This leads to a better purity of the signal and a smaller uncertainty on the background in the signal-enriched region.
The overall uncertainty on the cross section is slightly smaller than the result in Ref.~\cite{TOPQ-2018-05}, mainly because of an updated treatment of the systematic uncertainty for signal modelling.
 
\begin{figure}
\centering
\includegraphics[width=0.5\linewidth]{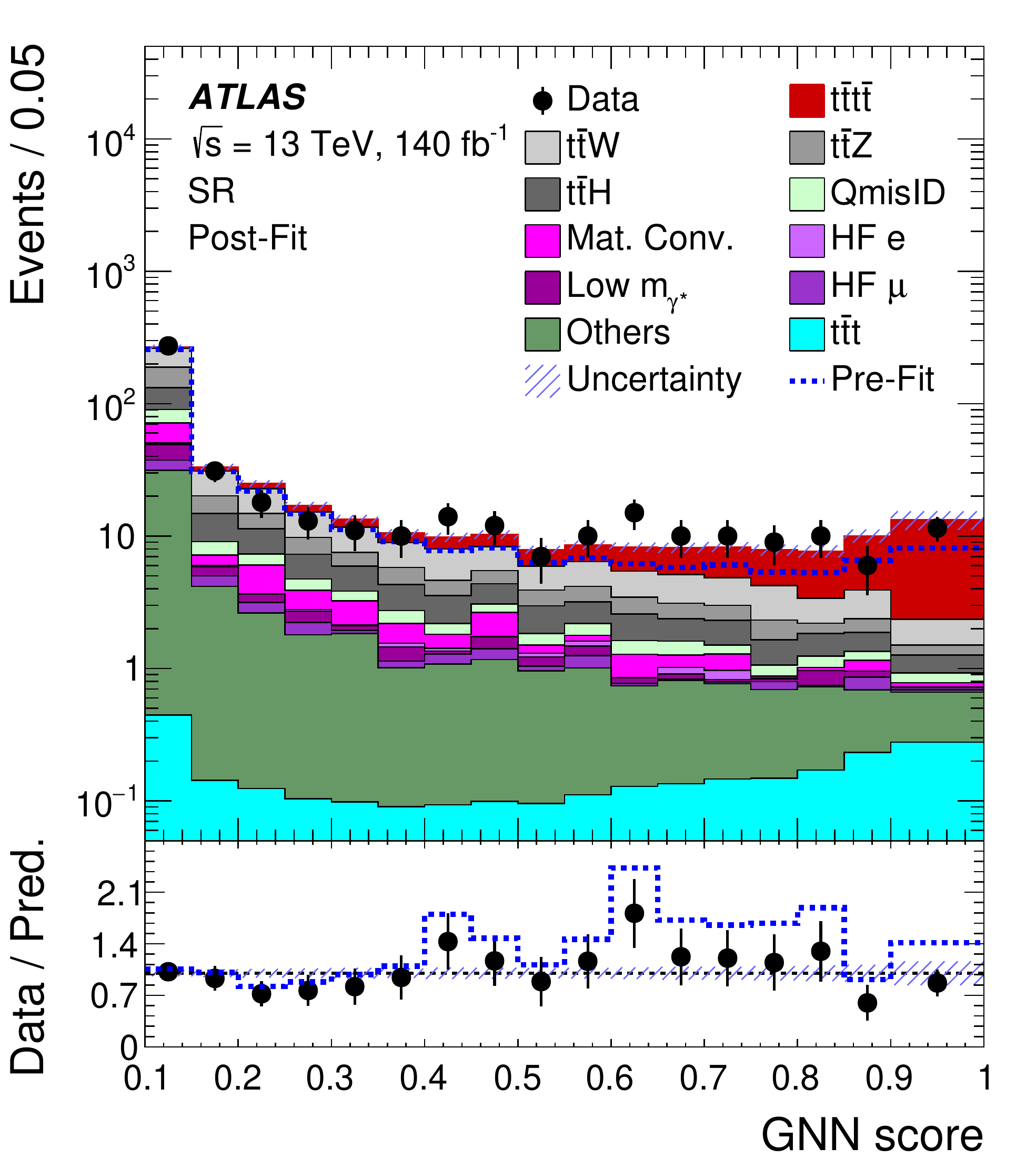}
\caption{\label{fig:SR_GNN} Comparison between data and the predictions after a fit to data for the GNN distribution in the SR.
The first bin contains underflow events. The ratio of the data to the total post-fit prediction is shown in the lower panel.
The dashed blue lines show the pre-fit prediction in the upper panel and the ratio of the data to the total pre-fit prediction in the lower panel.
The shaded band represents the total post-fit uncertainty in the prediction.
}
\end{figure}
 
The normalisation factors of the different fake/non-prompt lepton background sources and the parameters of the data-driven \ttW background model determined from the fit are shown in Tables~\ref{tab:CRfit} and \ref{tab:CRfit_ttW}.
The post-fit values of the background and signal yields before and after the fit as well as for events with a GNN score equal to or higher than 0.6 are shown in Table~\ref{tab:ssml:yields}.
The post-fit number of \ttW events with 6 and 7 jets is smaller than at the pre-fit level, while the number of \ttW events with $\geq$9 jets is increased compared to the pre-fit prediction. The overall number of fitted \ttW is in agreement with the \ttW cross section measurement in Ref.~\cite{TOPQ-2016-11}.
The number of background events from material conversion is also  increased.
The post-fit normalisation factors from Tables~\ref{tab:CRfit} and \ref{tab:CRfit_ttW} agree with their nominal value of 1, except for \NFCO.
They provide good agreement with data as shown in Table~\ref{tab:ssml:yields} and Figure~\ref{fig:SR_GNN}.

\begin{table}[!h]
\renewcommand*{\arraystretch}{1.6}
\centering
\caption{The normalisation factors for fake and non-prompt lepton background processes determined from the fit.
The uncertainties include both the statistical and systematic uncertainties.
The nominal pre-fit value for these factors is 1.}
\begin{tabular}{l|cccc}
\hline
Fake/non-prompt background & \NFCO &\NFGst & \NFHFe & \NFHFm\\
\hline
Value & $1.80^{+0.47}_{-0.41}$ & $1.08 ^{+0.37}_{-0.31}$ & $0.66^{+0.75}_{-0.46} $ & $1.27^{+0.53}_{-0.46}$ \\
\hline
\end{tabular}
\label{tab:CRfit}
\end{table}


\begin{table}[!h]
\renewcommand*{\arraystretch}{1.6}
\centering
\caption{The \ttW modelling parameters determined from the fit.
The uncertainties include both the statistical and systematic uncertainties.
The nominal pre-fit value for $a_0$ and $a_1$ is 0, while the nominal pre-fit value for \NFttWp and \NFttWm is 1.}
\begin{tabular}{l|cccc}
\hline
\ttW background & $a_0$ & $a_1$ & \NFttWp   & \NFttWm \\
\hline
Value & $0.51 \pm 0.10$ & $0.22^{+0.25}_{-0.22}$ & $1.27^{+0.25}_{-0.22}$ & $1.11^{+0.31}_{-0.28} $ \\
\hline
\end{tabular}
\label{tab:CRfit_ttW}
\end{table}


\begin{table}[htbp]
\renewcommand*{\arraystretch}{1.4}
\begin{center}
\caption{Pre-fit and post-fit background and signal yields in the signal region and for events with  GNN score larger than 0.6. The total systematic uncertainty differs from the sum in quadrature of the different uncertainties due to correlations.
\label{tab:ssml:yields}}
\begin{tabular}{lcccc}
\hline
&\multicolumn{2}{c}{Pre-fit}& \multicolumn{2}{c}{Post-fit}\\
\hline
& SR  & GNN$\geq$0.6 & SR  &  GNN$\geq$0.6\\
\hline
$t\bar{t}W$   & 130 $\pm$ 40 & 9 $\pm$ 4   & 127 $\pm$ 35& 12 $\pm$ 4\\
$t\bar{t}Z$   & 72 $\pm$ 15   & 3.4 $\pm$ 1.8& 79 $\pm$ 15 & 4.4 $\pm$ 2.0\\
$t\bar{t}H$   & 65 $\pm$ 11  & 4.6 $\pm$ 1.3 & 68 $\pm$ 10  & 5.0 $\pm$ 1.4 \\
\Qmis   & 27 $\pm$ 4  & 1.78 $\pm$ 0.26& 27 $\pm$ 4  & 1.80 $\pm$ 0.24\\
Mat. Conv.   & 16.5 $\pm$ 2.3& 0.73 $\pm$ 0.25  &30 $\pm$ 8 & 1.4 $\pm$ 0.5\\
HF e   & 3.1 $\pm$ 1.0 & 0.4 $\pm$ 0.5 & 2.3 $\pm$ 2.4& 0.3 $\pm$ 0.4 \\
HF $\mu$   & 7.1 $\pm$ 1.2 & 0.31 $\pm$ 0.15 & 9 $\pm$ 4& 0.41 $\pm$ 0.22 \\
Low $m_{\gamma^{*}}$   & 14.1 $\pm$ 2.0& 0.52 $\pm$ 0.19& 15 $\pm$ 5& 0.56 $\pm$ 0.22 \\
Others   & 47 $\pm$ 11 & 3.9 $\pm$ 1.2& 50 $\pm$ 10& 4.3 $\pm$ 1.2 \\
$t\bar{t}t$   & 2.9 $\pm$ 0.9 & 1.5 $\pm$ 0.5& 2.9 $\pm$ 0.9& 1.5 $\pm$ 0.5 \\
\hline
Total bkg & 390 $\pm$ 50 & 26 $\pm$ 5&  412 $\pm$ 21 & 32 $\pm$ 4 \\
\hline
$t\bar{t}t\bar{t}$   & 38 $\pm$ 4 & 25.2 $\pm$ 3.2&   69 $\pm$ 15 & 45 $\pm$ 10 \\
\hline
Total  & 430 $\pm$ 50 & 51 $\pm$ 7 & 480 $\pm$ 19 & 77 $\pm$ 8  \\
\hline
Data   & 482 & 83& 482& 83 \\
\hline
\end{tabular}
\end{center}
\end{table}


Figure~\ref{fig:SR_GNNhigh} presents the distributions of the number of jets, the number of $b$-jets, the sum of the four highest PCBT scores of jets in the event and $H_{T}$,
in the signal-enriched region for events
with a GNN score equal to or higher than 0.6. Good agreement between data and the post-fit predictions is observed.
 
\begin{figure}[htbp]
\centering
\subfloat[]{\includegraphics[width=0.4\linewidth]{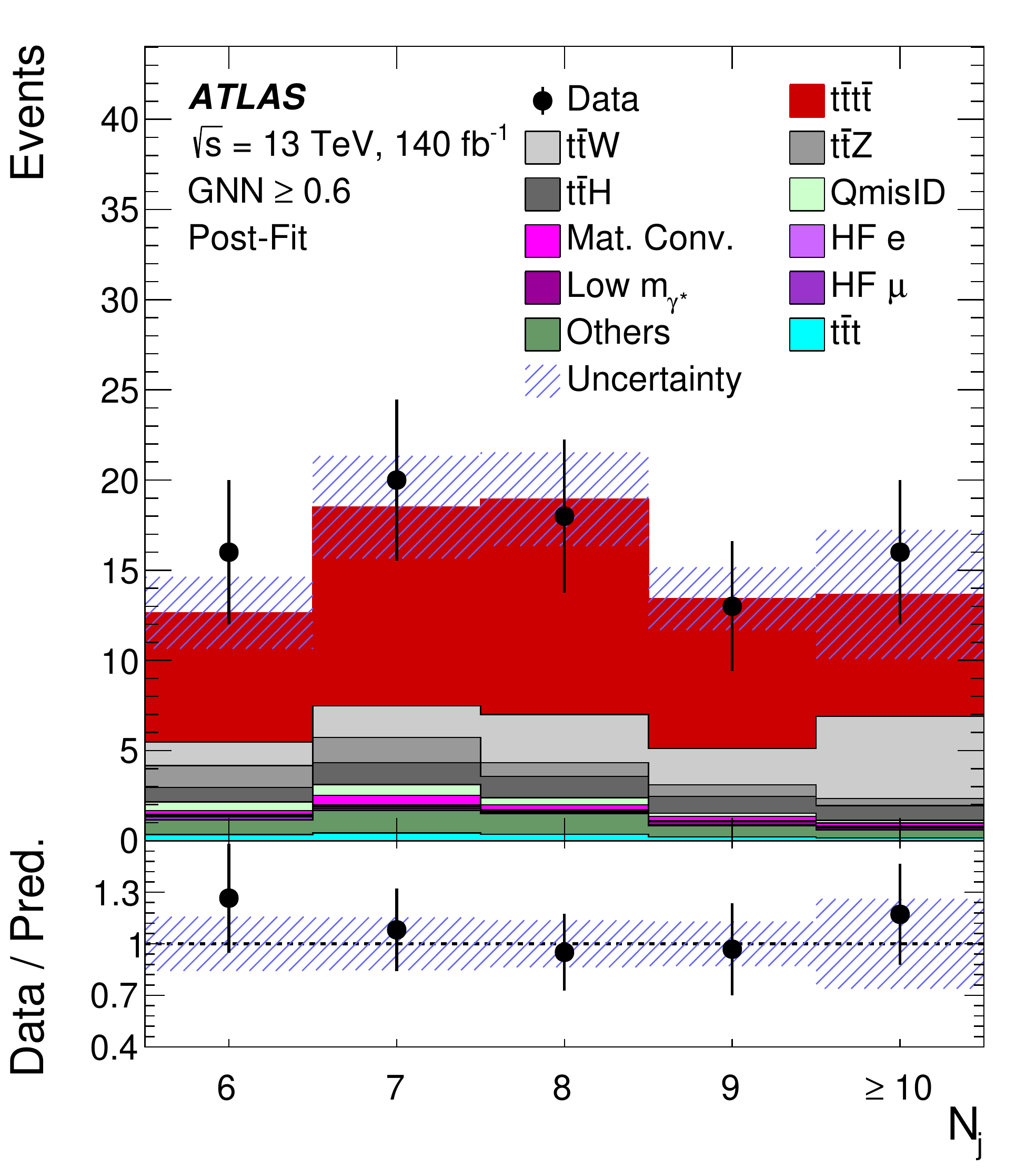}
}
\subfloat[]{\includegraphics[width=0.4\linewidth]{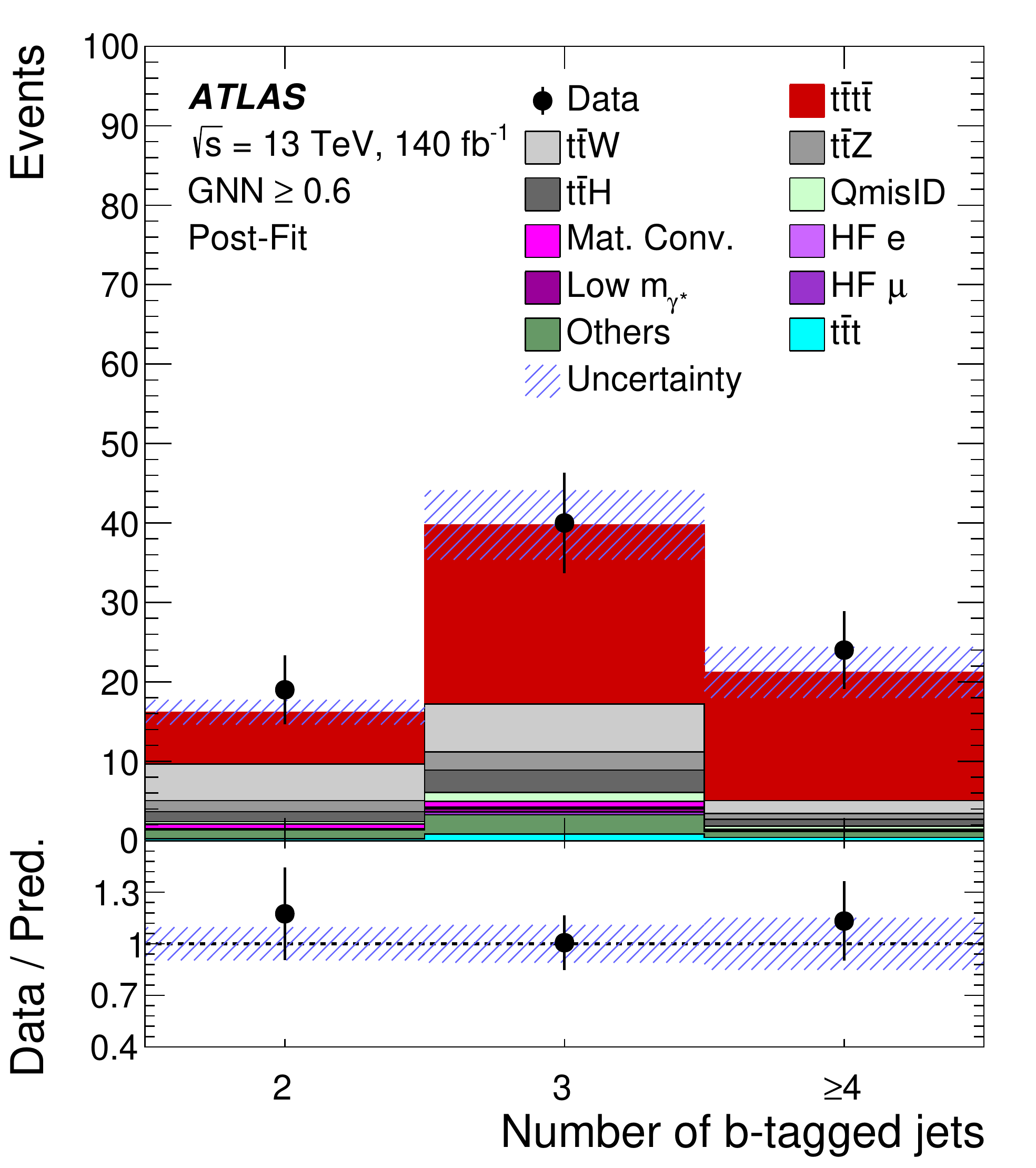}
}\\
\subfloat[]{\includegraphics[width=0.4\linewidth]{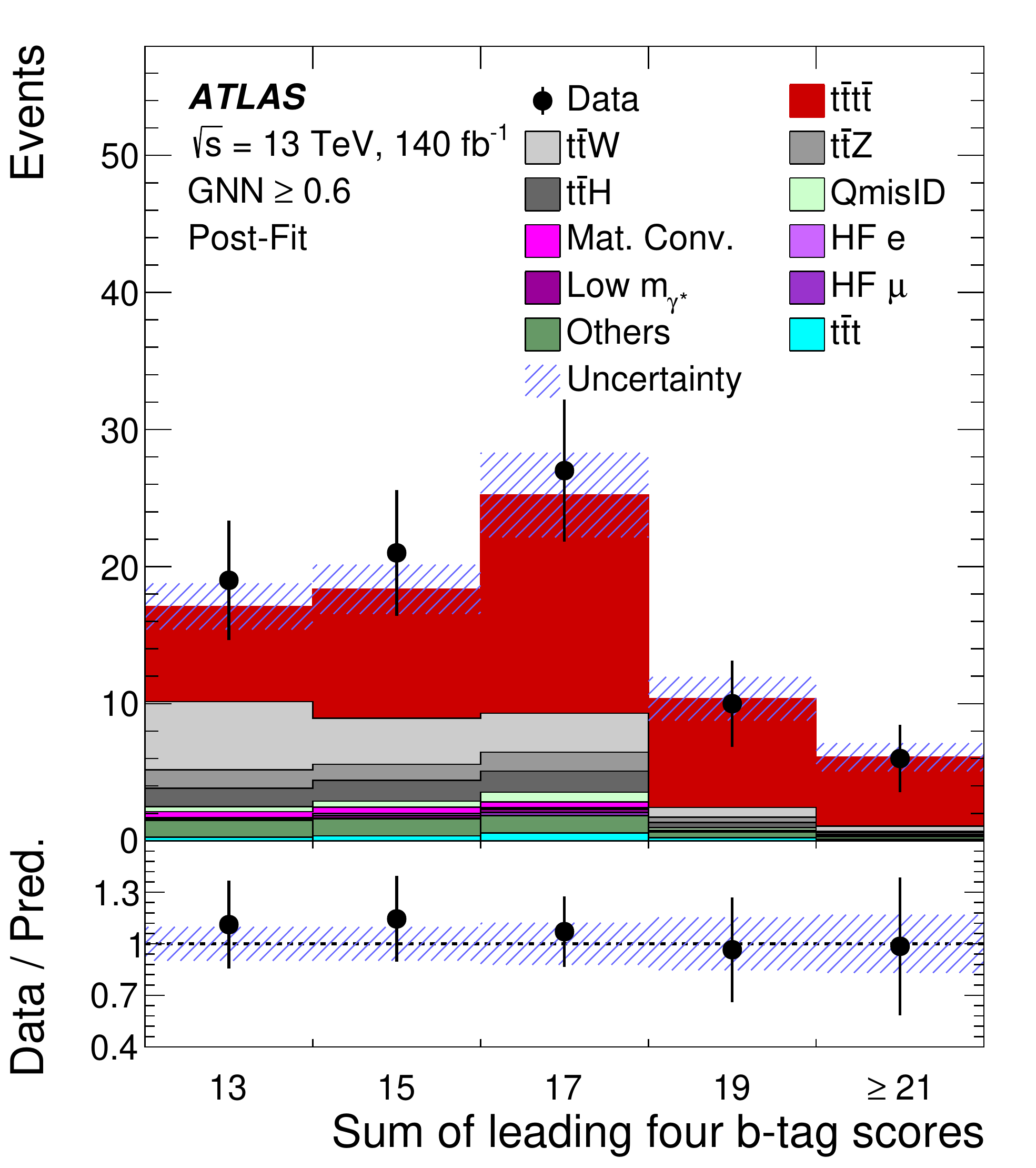}
}
\subfloat[]{\includegraphics[width=0.4\linewidth]{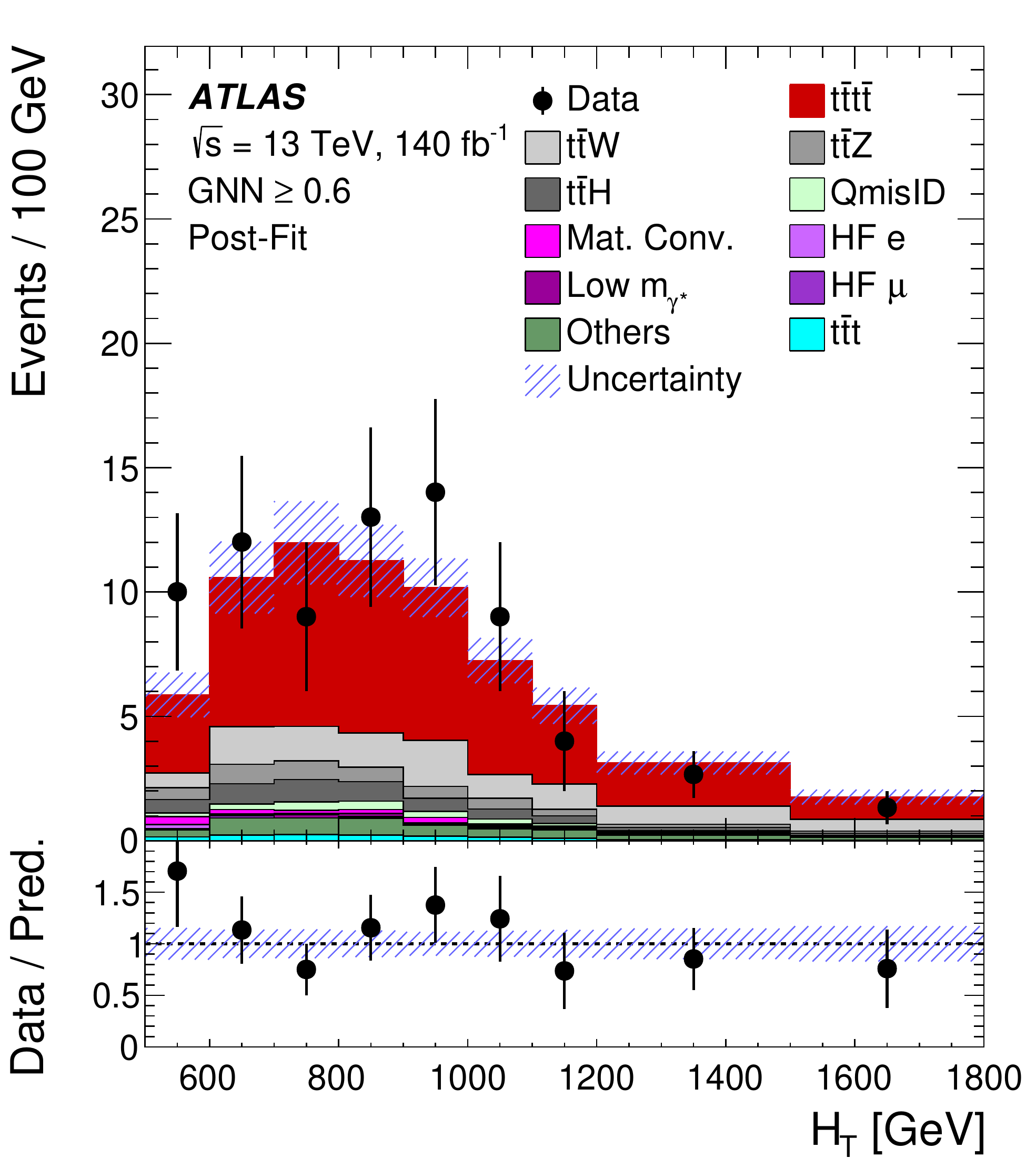}
}
\caption{\label{fig:SR_GNNhigh} Comparison between data and prediction after the fit to data in the signal-enriched region with GNN$\geq 0.6$ for the distributions of (a) the number of jets, (b) the number of $b$-jets, (c) the sum of the four highest PCBT scores of jets in the event, and (d) the sum of transverse momenta over all jets and leptons in the event ($H_{T}$). The ratio of the data to the total post-fit computation is shown in the lower panel. The shaded band represents the total post-fit uncertainty in the prediction. The first and last bins contain underflow and overflow events, respectively. }
\end{figure}
 
As a cross-check, a maximum-likelihood fit to the BDT score distribution in data is also performed yielding the same signal strength corresponding to the
observed significance of 6.0 standard deviations while 3.9 is expected.
 
The largest systematic uncertainties in the measurement of $\sigma_{\tttt}$ arise from the modelling
of the \tttt signal, with the uncertainty from the choice of the MC generator being by far
the largest followed by the uncertainty from the parton shower and hadronisation model.
The uncertainty in the data-driven \ttW background estimate is also significant. Among the experimental uncertainties, the largest effects come from
the $b$-jet tagging and the jet energy scale.
The parameter associated with the \tttt MC generator choice is pulled by around $-$0.5 standard deviations after the fit. No other nuisance parameter is found to be significantly adjusted or constrained by the fit.
The uncertainties impacting the \tttt cross section measurement are summarised in Table~\ref{tab:syst}.

\begin{table}[htbp]
\centering
\caption{List of the uncertainties in the cross section $\sigma_{\tttt}$ , grouped in categories. The quoted values are obtained by repeating the fit, fixing a set of nuisance parameters of the sources corresponding to each category, and subtracting in quadrature the resulting uncertainty from the total uncertainty of the nominal fit presented in the last row. The total uncertainty is different from the sum in quadrature of the components due to correlations among nuisance parameters.
\label{tab:syst}}
\begin{tabular}{l  c  c  c c }
\hline
Uncertainty source & \multicolumn{2}{c}{$\Delta\sigma$ [fb]} & \multicolumn{2}{c}{ $\Delta\sigma/\sigma$[\%]}  \\
\hline
\textbf{Signal modelling} & \multicolumn{4}{c}{}   \\
$t\bar{t}t\bar{t}$ generator choice &  +3.7 & $-$2.7  &  +17 & $-$12 \\
$t\bar{t}t\bar{t}$ parton shower model & +1.6 & $-$1.0 & +7 & $-$4 \\
Other $t\bar{t}t\bar{t}$ modelling & +0.8 & $-$0.5 & +4 & $-$2 \\
\hline
\textbf{Background modelling} & \multicolumn{4}{c}{}  \\
$t\bar{t}H$+jets modelling  & +0.9 & $-$0.7 & +4 & $-$3 \\
$t\bar{t}W$+jets modelling & +0.8 & $-$0.8 & +4 & $-$3 \\
$t\bar{t}Z$+jets modelling   & +0.5 & $-$0.4 & +2 & $-$2 \\
Other background modelling & +0.5 & $-$0.4 & +2 & $-$2 \\
Non-prompt leptons modelling & +0.4 & $-$0.3 & +2 & $-$2 \\
$t\bar{t}t$ modelling & +0.3 & $-$0.2 & +1 & $-$1 \\
Charge misassignment & +0.1 & $-$0.1 & +0 & $-$0 \\
\hline
\textbf{Instrumental} & \multicolumn{4}{c}{}  \\
Jet flavour tagging ($b$-jets) & +1.1 & $-$0.8 & +5 & $-$4 \\
Jet uncertainties & +1.1 & $-$0.7 & +5 & $-$3 \\
Jet flavour tagging (light-flavour jets) & +0.9 & $-$0.6 & +4 & $-$3 \\
Jet flavour tagging ($c$-jets) & +0.5 & $-$0.4 & +2 & $-$2 \\
Simulation sample size & +0.4 & $-$0.3 & +2 & $-$1 \\
Other experimental uncertainties & +0.4 & $-$0.3 & +2 & $-$1 \\
Luminosity & +0.2 & $-$0.2 & +1 & $-$1 \\
\hline
Total systematic uncertainty & +4.6 & $-$3.4 & +20 & $-$16 \\
\hline
\textbf{Statistical} &  & &  &  \\
Intrinsic statistical uncertainty & +4.2 &$-$3.9  & +19 & $-$17\\
$t\bar{t}W$+jets normalisation and scaling factors & +1.2 & $-$1.1 & +6 & $-$5 \\
Non-prompt leptons normalisation (HF, Mat. Conv., Low $m_{\gamma^{*}}$) & +0.4 & $-$0.3 & +2 & $-$1 \\
\hline
Total statistical uncertainty &  +4.7 & $-$4.3 & +21 & $-$19 \\
\hline
Total uncertainty  & +6.6 & $-$5.5 & +29 & $-$25 \\
\hline
\end{tabular}
\end{table}


This analysis derives 95\% confidence level (CL) intervals on the cross section of the \ttt process, $\sigma_{\ttt}$.
In the SM, triple top quarks  are always produced in association with other particles, and can be split into the \tttW\ and \tttq processes~\cite{Barger:2010uw,Chen:2014,Boos:2021yat}. These processes give rise to an experimental signature and kinematic properties of the event similar to that of the \tttt signal, resulting in a similar GNN shape for \tttt and \ttt production.
Varying the normalisation of the \tttt and \ttt processes in the likelihood fit simultaneously results in an anti-correlation between the two processes of $-$93\%.
Figure~\ref{fig:3top_4top:2D} shows the two-dimensional negative log-likelihood contour for the likelihood fit where the \ttt\ and \tttt\ cross sections are treated as free parameters.
Although the best-fit value in this case is consistent with a \tttt cross section of zero, it is also consistent with the SM prediction at the level of 2.1 standard deviations.
 
The two components of the \ttt process (\tttW and \tttq) have different sensitivity to possible BSM effects. In particular, a number of higher-dimensional operators generating flavour changing neutral currents involving the top quark~\cite{Cao:2019qrb,Khanpour:2019qnw} lead to the production of \ttt events without an associated $W$ boson in the final state. Thus it is interesting to derive 95\% CL intervals on the cross sections of the \tttW\ and \tttq\ processes separately. Under the assumption of a \tttt signal strength of 1 and of 1.9, the 95\% CL intervals on \ttt, \tttW\ and \tttq\ are shown in Table~\ref{tab:ttt_results}.
The 95\% CL intervals for the \tttq\ process are wider than those for the \tttW\ process because the \tttq\ process is characterised by lower lepton and jet multiplicities in the final state, which results in a lower selection efficiency.
 
\begin{figure}
\centering
\includegraphics[width=0.60\linewidth]{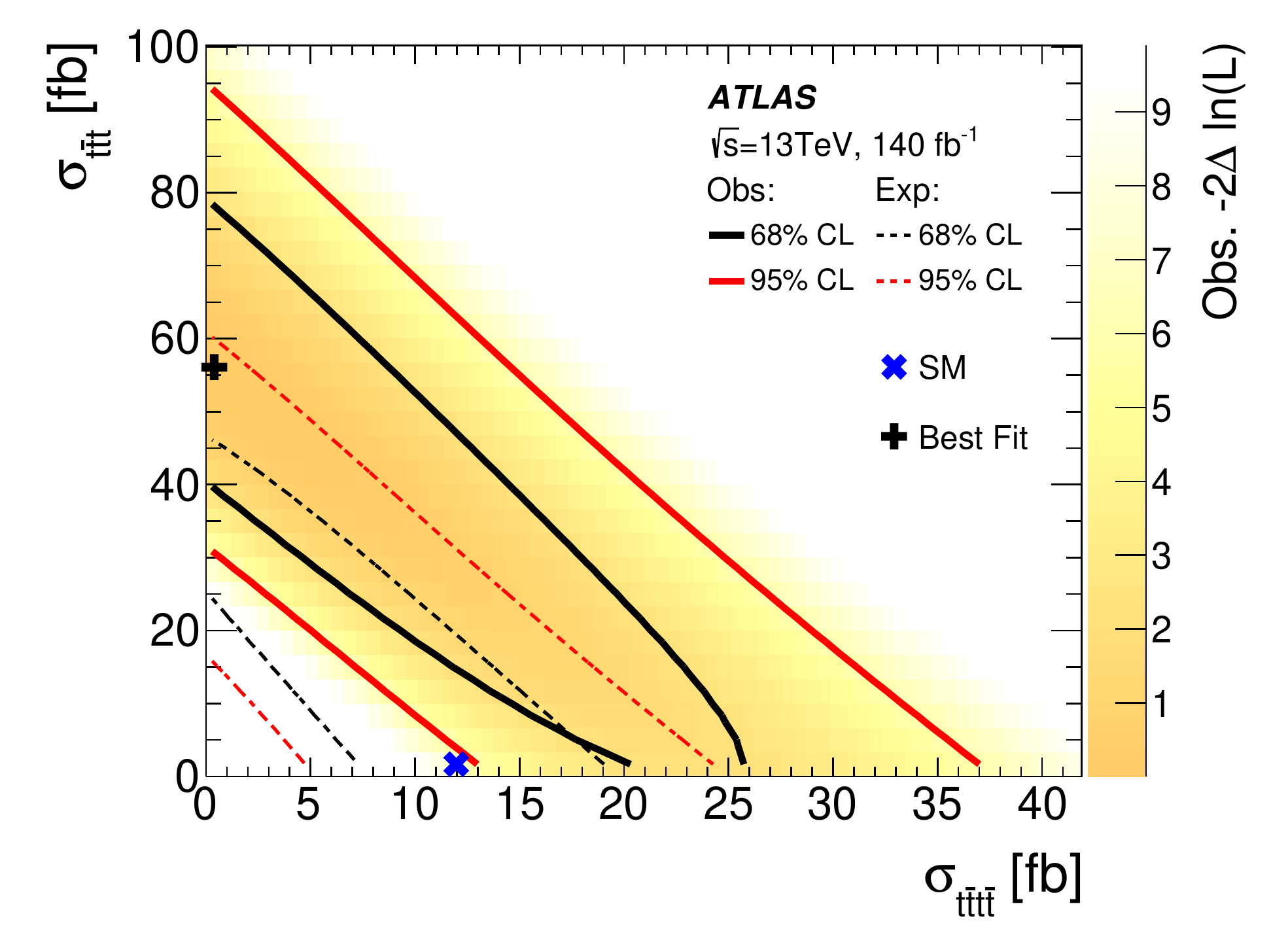}
\caption{\label{fig:3top_4top:2D} Two-dimensional negative log-likelihood contour for the $\ttt$ cross section ($\sigma_{\ttt}$) versus the $\tttt$ cross section ($\sigma_{\tttt}$) when the normalisation of both processes are treated as free parameters in the fit.
The blue cross shows the SM expectation of $\stttt=12$~fb from Ref.~\cite{Frederix:2017wme} and $\sigma_{\ttt}=1.67$~fb (see Section~\ref{sec:mcsamples}), both computed at NLO, while the black cross shows the best-fit value.
The observed (expected) exclusion contours at 68\% (black) and 95\% CL (red) are shown in solid (dashed) lines.
The gradient-shaded area represents the observed likelihood value as a function of $\sigma_{\ttt}$ and $\sigma_{\tttt}$.
}
\end{figure}

\begin{table}[htbp!]
\renewcommand*{\arraystretch}{1.1}
\centering
{\setlength\tabcolsep{12.0pt}
\caption{Observed 95\% CL intervals for the \ttt, \tttq and \tttW cross sections assuming a \tttt signal strength of 1 and 1.9. To derive the \tttW(\tttq) cross section interval, the \tttq(\tttW) cross section is fixed to its SM prediction.
}
\begin{tabular}{c| c c}
\toprule
Processes & \multicolumn{2}{c}{95\% CL cross section interval [fb]}  \\
& $\mu_{\tttt} = 1$ & $\mu_{\tttt} = 1.9$ \\
\midrule
\ttt  & [4.7, 60] & [0, 41]\\
\tttW & [3.1, 43] & [0, 30]\\
\tttq & [0, 144]  & [0, 100]\\
\bottomrule
\end{tabular}
\label{tab:ttt_results}
}
\end{table}



\section{Interpretations}
Limits and 95\% CL intervals are also
obtained on the top-quark Yukawa coupling, on EFT operators that parametrize BSM \tttt\ production, and on a Higgs oblique parameter.
 
\subsection{Limits on the top-quark Yukawa coupling}
The top-quark Yukawa coupling enters the electroweak \tttt Feynman diagram where a pair of top quarks is mediated by a Higgs boson (see Figure~\ref{fig:feynman}). This makes \tttt production sensitive to the top-quark Yukawa coupling, including the coupling strength and its $CP$ properties. An additional dependence on the Yukawa coupling comes through the \ttHiggs background.
The \tttt cross section can be parameterised as a function of two parameters: the top Yukawa coupling strength modifier $\kappa_t$, and the $CP$-mixing angle $\alpha$~\cite{Cao:2019ygh, HIGG-2019-01}.
The SM corresponds to a pure $CP$-even coupling with $\alpha$ = 0 and  $\kappa_t$ = 1, while a $CP$-odd coupling is realised when $\alpha$ = 90$^{\circ}$.
In case of a pure $CP$-even coupling, $\kappa_t$ can be expressed as a ratio of the top-quark Yukawa coupling $y_t$ to the SM prediction $y_t^\mathrm{SM}$.
The \ttHiggs cross section also depends on these parameters, but unlike \tttt production, the \ttHiggs kinematic distributions change only when the $CP$-odd term $\kappa_t\sin(\alpha)$ is non-zero. The \tttt and \ttHiggs yields in each bin of the GNN distribution are parameterised as a function of $\kappa_t$ and $\alpha$.
The observed (expected) 95\% CL limits are shown in Figure~\ref{fig:yukawa} in the two-dimensional parameter space
($|\kappa_t\cos(\alpha)|, |\kappa_t\sin(\alpha)|$).
Fixing the top-quark Yukawa coupling to be $CP$-even only (i.e. $\alpha =0$),
the following observed (expected) limits are extracted: $|\kappa_t| < 1.9$ (1.6).
This limit is less stringent than that reported in Ref.~\cite{CMS-TOP-18-003} using a similar technique, due to a slightly higher measured SM \tttt\ cross section than the prediction.
As a check to probe the effect from the \ttHiggs process, an alternative fit is performed. As opposed to the fit used to obtain the nominal result, the \ttHiggs background yields are not parametrised, whilst the normalisation of the \ttHiggs background is treated as a free parameter of the fit. This leads to an observed (expected) limit of $|\kappa_t| < 2.3$ (1.9).
\begin{figure}
\centering
\includegraphics[width=0.60\linewidth]{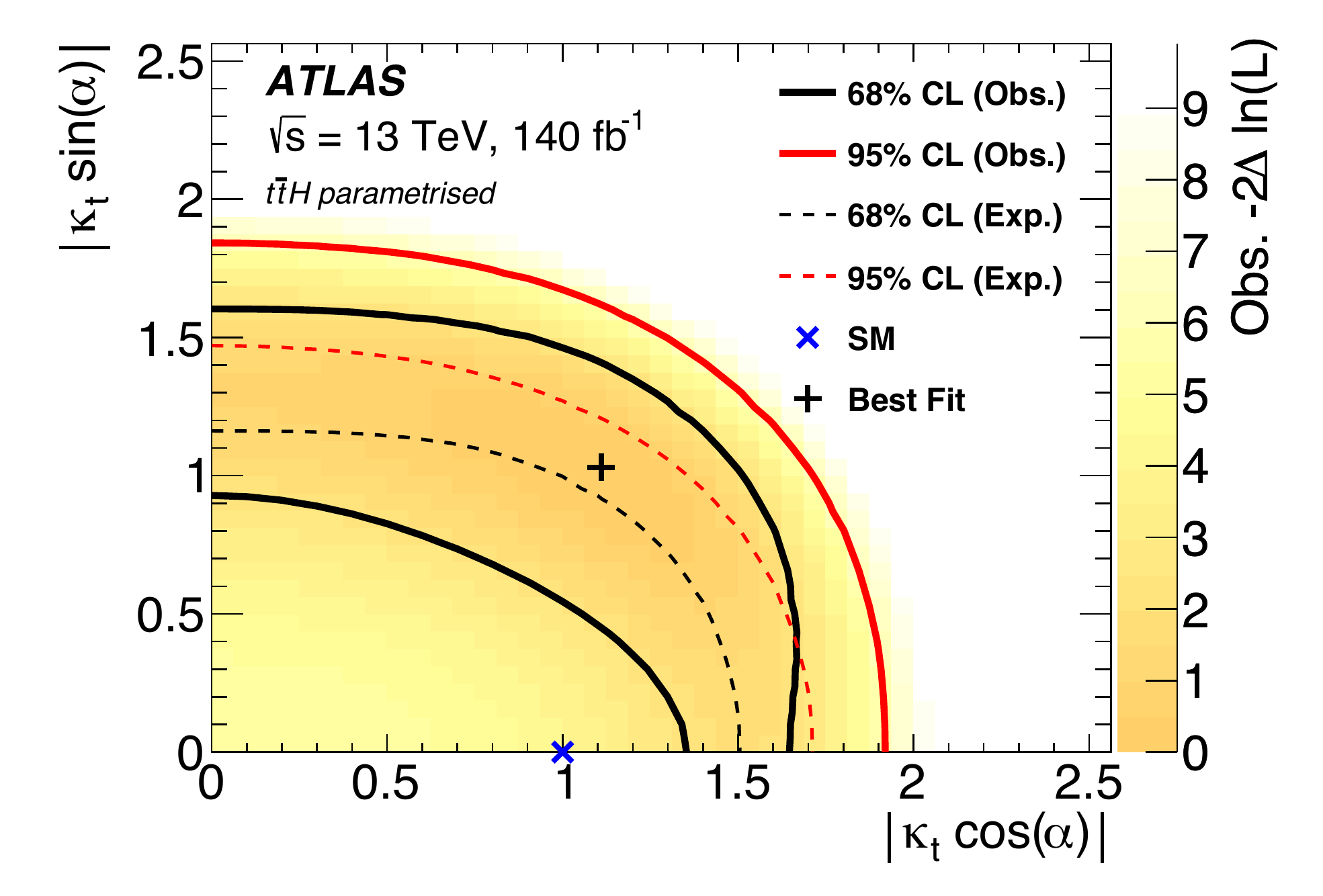}
\caption{\label{fig:yukawa} Two-dimensional negative log-likelihood contours for |$\kappa_t\cos(\alpha)$| versus |$\kappa_t\sin(\alpha)$| at 68\% and 95\%, where $\kappa_t$ is the top-Higgs Yukawa coupling strength parameter and $\alpha$ is the mixing angle between the $CP$-even and $CP$-odd components.
The gradient-shaded area represents the observed likelihood value as a function of $\kappa_t$ and $\alpha$.
Both the \tttt signal and  \ttHiggs background yields in each fitted bin are parameterised as a function of $\kappa_t$ and $\alpha$. The blue cross shows the SM expectation, while the black cross shows the best fit value.
}
\end{figure}
 
\subsection{Limits on EFT operators and the Higgs oblique parameter}
 
Within the EFT framework, the \tttt process is sensitive to four heavy-flavour fermion operators $O_{tt}^1$, $O_{QQ}^1$, $O_{Qt}^1$ and $O_{Qt}^8$, which can probe the BSM models that enhance interactions between the third-generation quarks~\cite{Zhang:2017mls}.
The \tttt production cross section can be approximated by:
\begin{equation}{\sigma}_{t\Bar{t}t\Bar{t}} = {\sigma}_{t\Bar{t}t\Bar{t}}^{SM} + \frac{1}{\Lambda^2}\sum_{i}C_{i}{\sigma}_{i}^{(1)}+
\frac{1}{\Lambda^{4}}\sum_{i\leq j}C_{i}C_{j}{\sigma}_{i,j}^{(2)} ,
\label{equation:EFT_par}
\end{equation}
where $C_{i}$ denotes the coupling parameters of the four heavy-flavour fermion operators, $C_{i}{\sigma}_{i}^{(1)} $ is the linear term that represents the interference of dimension-6 operators with SM operators, and
$C_{i}C_{i}{\sigma}_{i,j}^{(2)}$ is the quadratic term that also includes the interference between different EFT operators.
The 95\% CL intervals on the EFT parameters are extracted by parameterising the \tttt yield in each bin of the GNN score distribution as a quadratic function of the coefficient of the corresponding EFT operator ($C_i/\Lambda^2$) and performing the fit to data. The fit is carried out assuming that only one operator contributes to the \tttt cross section, while the coefficients of the other three operators are fixed to the SM value of zero. The expected and observed 95\% CL intervals on the coefficients are summarised in Table~\ref{tab:EFTresults}.
To probe the importance of the different terms in Equation~\ref{equation:EFT_par}, the limits are also extracted assuming only the linear terms as a test.
The resulting upper limits on the absolute values of the coefficients ($|C_i/\Lambda^2|$) of $O_{QQ}^1$, $O_{Qt}^1$, $O_{tt}^1$ and $O_{Qt}^8$ are
6.6, 4.0, 2.8 and 10.8 \TeV$^{-2}$, respectively, at 95\%CL. Comparable limits on these EFT parameters can be found in Ref.~\cite{CMS-TOP-17-019}.

\begin{table}[htbp!]
\centering
\caption{Expected and observed 95\% CL intervals on EFT coupling parameters assuming one EFT parameter variation in the fit.}
\begin{tabular}{c| c c}
\toprule
Operators & Expected ${C_{i}}/{\Lambda^2}$ [\TeV$^{-2}$] & Observed  ${C_{i}}/{\Lambda^2}$ [\TeV$^{-2}$] \\
\midrule
$\mathcal{O}^{1}_{QQ}$ & [-2.5, 3.2] & [-4.0, 4.5]\\
$\mathcal{O}^{1}_{Qt}$ & [-2.6, 2.1] & [-3.8, 3.4]\\
$\mathcal{O}^{1}_{tt}$ & [-1.2, 1.4] & [-1.9, 2.1]\\
$\mathcal{O}^{8}_{Qt}$ & [-4.3, 5.1] & [-6.9, 7.6] \\
\bottomrule
\end{tabular}
\label{tab:EFTresults}
\end{table}


An oblique parameter is a self-energy correction term applied to electroweak propagators in the SM. The BSM additions to such a correction can be expressed as an EFT expansion, with the self-energy correction to the Higgs boson parameterised by the parameter $\hat{H}$, where $\hat{H}=0$ corresponds to the SM prediction~\cite{Englert:2019zmt}. The $\hat{H}$ parameter affects the off-shell Higgs interaction, and thus the \tttt cross section, as well as processes involving a Higgs boson, in particular \ttHiggs production, which is a significant background to the \tttt measurement. To account for this effect,
the \ttHiggs contribution is parameterised as a function of $\hat{H}$~\cite{Englert:2019zmt}: $\mu_{\ttHiggs} = 1-\hat{H}$, where $\mu_{\ttHiggs}$ is the \ttHiggs normalisation factor with respect to the SM cross section. A limit on $\hat{H}$ is extracted from the likelihood scans shown on Figure~\ref{fig:higgsoblique}. The observed (expected) upper limit on the $\hat{H}$ value is 0.23 (0.11) at  95\% CL. The observed limit is weaker than the largest value of this parameter equal to 0.2 that preserves unitarity in the perturbative theory. Previously, limits on the $\hat{H}$ parameter were reported in Refs.~\cite{Englert:2019zmt,HIGG-2018-57,CMS-TOP-18-003}.
\begin{figure}
\centering
\includegraphics[width=0.60\linewidth]{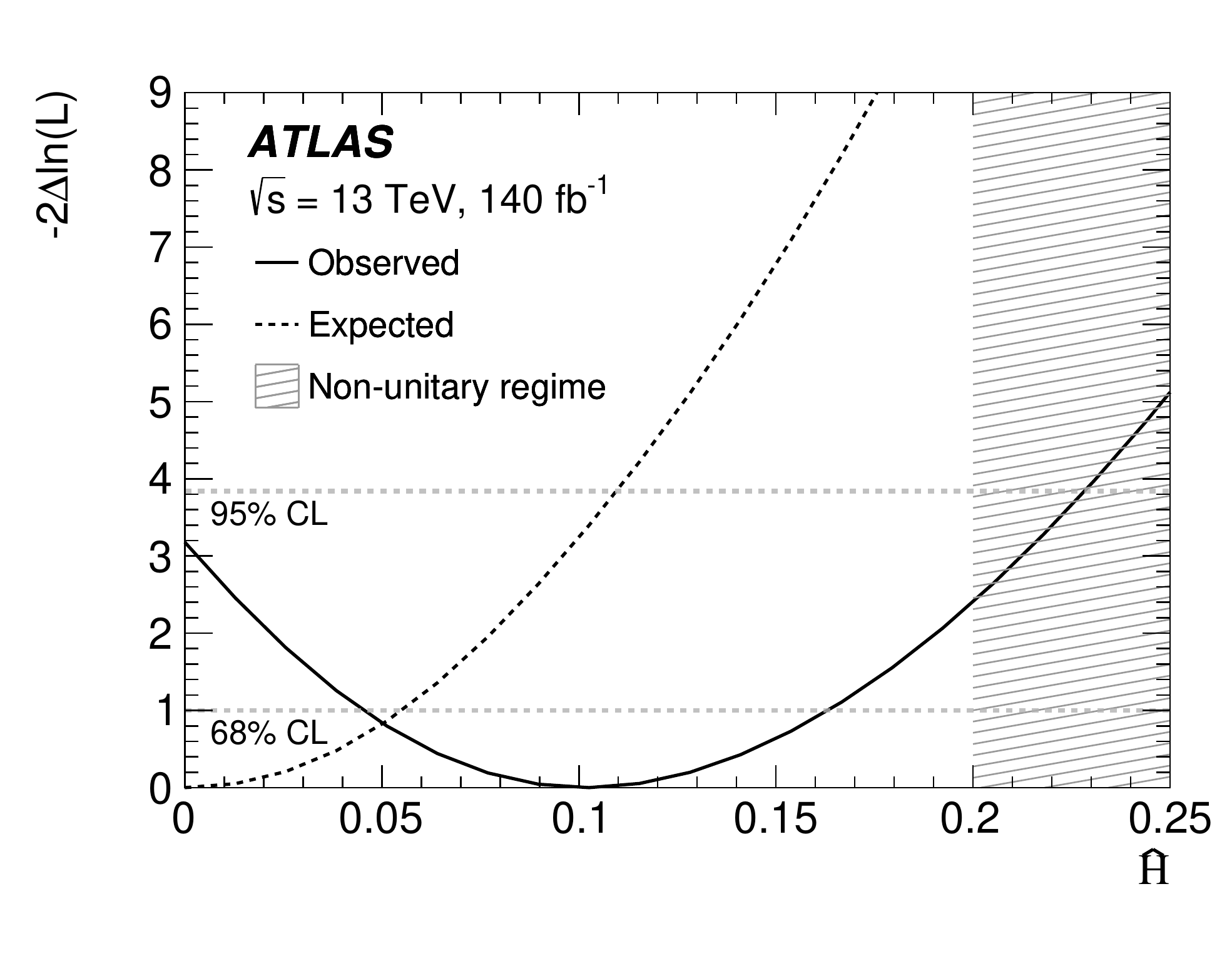}
\caption{\label{fig:higgsoblique} The negative log-likelihood values as a function of the Higgs oblique parameter $\hat{H}$. The solid line represents the observed likelihood while the dashed line corresponds to the expected one. The dashed region shows the non-unitary regime.}
\end{figure}


\FloatBarrier

\section{Conclusion}
This paper presents a measurement of \tttt production using 140~\ifb\ of data at $\sqrt{s}=13~\TeV$ with the ATLAS detector at the LHC.
Events are selected with exactly two same-charge isolated leptons or at least three isolated leptons.
The normalisation of the \ttW background in jet multiplicity bins is determined using a data-driven approach.
A Graph Neural Network is used to separate the \tttt signal from the background.
Four-top production is observed with a significance of 6.1 standard deviations
with respect to the background-only hypothesis. The expected significance is 4.3 or 4.7 standard deviations
depending on the assumed SM cross section.
The measured \tttt\ production cross section is $22.5^{+6.6}_{-5.5}$~fb. It is
consistent with the SM predictions within 1.8  or 1.7 standard deviations and with the previous ATLAS measurement.
In addition, 95\% confidence level intervals on the cross section for inclusive \ttt\ production and its subprocesses \tttq\ and \tttW\ are provided, for the scenarios assuming the SM \tttt\ production cross section and the measured \tttt\ cross section.
 
The results are used to set limits on several new physics scenarios.
Constraints on the $CP$ properties of the top-quark Yukawa coupling are obtained in the form of limits in the two-dimensional parameter space $\left(|\kappa_t\cos(\alpha)|, |\kappa_t\sin(\alpha)|\right)$.
Assuming a pure $CP$-even coupling ($\alpha=0$), the observed upper limit on $|\kappa_t|=|y_t/y_t^\mathrm{SM}|$ at 95\% CL is 1.9.
Constraints at 95\% CL are obtained on the four dimension-6 heavy-flavour fermion operators.
Assuming one operator taking effect at a time,
the observed constraints on the coefficients ($C_i/\Lambda^2$) of $O_{QQ}^1$, $O_{Qt}^1$, $O_{tt}^1$ and $O_{Qt}^8$
are $[-4.0, 4.5]$, $[-3.8, 3.4]$, $[-1.9, 2.1]$ and $[-6.9, 7.6]$ \TeV$^{-2}$, respectively.
An observed upper limit at 95\% CL of 0.23 is obtained for the Higgs oblique parameter that is weaker than the largest value that preserves unitarity in the perturbative theory.


\section*{Acknowledgements}


We thank CERN for the very successful operation of the LHC, as well as the
support staff from our institutions without whom ATLAS could not be
operated efficiently.
 
We acknowledge the support of
ANPCyT, Argentina;
YerPhI, Armenia;
ARC, Australia;
BMWFW and FWF, Austria;
ANAS, Azerbaijan;
CNPq and FAPESP, Brazil;
NSERC, NRC and CFI, Canada;
CERN;
ANID, Chile;
CAS, MOST and NSFC, China;
Minciencias, Colombia;
MEYS CR, Czech Republic;
DNRF and DNSRC, Denmark;
IN2P3-CNRS and CEA-DRF/IRFU, France;
SRNSFG, Georgia;
BMBF, HGF and MPG, Germany;
GSRI, Greece;
RGC and Hong Kong SAR, China;
ISF and Benoziyo Center, Israel;
INFN, Italy;
MEXT and JSPS, Japan;
CNRST, Morocco;
NWO, Netherlands;
RCN, Norway;
MEiN, Poland;
FCT, Portugal;
MNE/IFA, Romania;
MESTD, Serbia;
MSSR, Slovakia;
ARRS and MIZ\v{S}, Slovenia;
DSI/NRF, South Africa;
MICINN, Spain;
SRC and Wallenberg Foundation, Sweden;
SERI, SNSF and Cantons of Bern and Geneva, Switzerland;
MOST, Taiwan;
TENMAK, T\"urkiye;
STFC, United Kingdom;
DOE and NSF, United States of America.
In addition, individual groups and members have received support from
BCKDF, CANARIE, Compute Canada and CRC, Canada;
PRIMUS 21/SCI/017 and UNCE SCI/013, Czech Republic;
COST, ERC, ERDF, Horizon 2020 and Marie Sk{\l}odowska-Curie Actions, European Union;
Investissements d'Avenir Labex, Investissements d'Avenir Idex and ANR, France;
DFG and AvH Foundation, Germany;
Herakleitos, Thales and Aristeia programmes co-financed by EU-ESF and the Greek NSRF, Greece;
BSF-NSF and MINERVA, Israel;
Norwegian Financial Mechanism 2014-2021, Norway;
NCN and NAWA, Poland;
La Caixa Banking Foundation, CERCA Programme Generalitat de Catalunya and PROMETEO and GenT Programmes Generalitat Valenciana, Spain;
G\"{o}ran Gustafssons Stiftelse, Sweden;
The Royal Society and Leverhulme Trust, United Kingdom.
 
The crucial computing support from all WLCG partners is acknowledged gratefully, in particular from CERN, the ATLAS Tier-1 facilities at TRIUMF (Canada), NDGF (Denmark, Norway, Sweden), CC-IN2P3 (France), KIT/GridKA (Germany), INFN-CNAF (Italy), NL-T1 (Netherlands), PIC (Spain), ASGC (Taiwan), RAL (UK) and BNL (USA), the Tier-2 facilities worldwide and large non-WLCG resource providers. Major contributors of computing resources are listed in Ref.~\cite{ATL-SOFT-PUB-2021-003}.


\clearpage
 
\printbibliography
 
\clearpage
 
\begin{flushleft}
\hypersetup{urlcolor=black}
{\Large The ATLAS Collaboration}

\bigskip

\AtlasOrcid[0000-0002-6665-4934]{G.~Aad}$^\textrm{\scriptsize 102}$,
\AtlasOrcid[0000-0002-5888-2734]{B.~Abbott}$^\textrm{\scriptsize 120}$,
\AtlasOrcid[0000-0002-1002-1652]{K.~Abeling}$^\textrm{\scriptsize 55}$,
\AtlasOrcid[0000-0001-5763-2760]{N.J.~Abicht}$^\textrm{\scriptsize 49}$,
\AtlasOrcid[0000-0002-8496-9294]{S.H.~Abidi}$^\textrm{\scriptsize 29}$,
\AtlasOrcid[0000-0002-9987-2292]{A.~Aboulhorma}$^\textrm{\scriptsize 35e}$,
\AtlasOrcid[0000-0001-5329-6640]{H.~Abramowicz}$^\textrm{\scriptsize 151}$,
\AtlasOrcid[0000-0002-1599-2896]{H.~Abreu}$^\textrm{\scriptsize 150}$,
\AtlasOrcid[0000-0003-0403-3697]{Y.~Abulaiti}$^\textrm{\scriptsize 117}$,
\AtlasOrcid[0000-0003-0762-7204]{A.C.~Abusleme~Hoffman}$^\textrm{\scriptsize 137a}$
% ...
\bigskip
\\

% ...
$^{29}$Physics Department, Brookhaven National Laboratory, Upton NY; United States of America.\\
% ...
$^{35}$$^{(a)}$Facult\'e des Sciences Ain Chock, R\'eseau Universitaire de Physique des Hautes Energies - Universit\'e Hassan II, Casablanca;$^{(b)}$Facult\'{e} des Sciences, Universit\'{e} Ibn-Tofail, K\'{e}nitra;$^{(c)}$Facult\'e des Sciences Semlalia, Universit\'e Cadi Ayyad, LPHEA-Marrakech;$^{(d)}$LPMR, Facult\'e des Sciences, Universit\'e Mohamed Premier, Oujda;$^{(e)}$Facult\'e des sciences, Universit\'e Mohammed V, Rabat;$^{(f)}$Institute of Applied Physics, Mohammed VI Polytechnic University, Ben Guerir; Morocco.\\
% ...
$^{49}$Fakult\"{a}t Physik , Technische Universit{\"a}t Dortmund, Dortmund; Germany.\\
% ...
$^{55}$II. Physikalisches Institut, Georg-August-Universit\"{a}t G\"ottingen, G\"ottingen; Germany.\\
% ...
$^{102}$CPPM, Aix-Marseille Universit\'e, CNRS/IN2P3, Marseille; France.\\
% ...
$^{117}$Department of Physics, New York University, New York NY; United States of America.\\
% ...
$^{120}$Homer L. Dodge Department of Physics and Astronomy, University of Oklahoma, Norman OK; United States of America.\\
% ...
$^{137}$$^{(a)}$Departamento de F\'isica, Pontificia Universidad Cat\'olica de Chile, Santiago;$^{(b)}$Millennium Institute for Subatomic physics at high energy frontier (SAPHIR), Santiago;$^{(c)}$Instituto de Investigaci\'on Multidisciplinario en Ciencia y Tecnolog\'ia, y Departamento de F\'isica, Universidad de La Serena;$^{(d)}$Universidad Andres Bello, Department of Physics, Santiago;$^{(e)}$Instituto de Alta Investigaci\'on, Universidad de Tarapac\'a, Arica;$^{(f)}$Departamento de F\'isica, Universidad T\'ecnica Federico Santa Mar\'ia, Valpara\'iso; Chile.\\
% ...
$^{150}$Department of Physics, Technion, Israel Institute of Technology, Haifa; Israel.\\
$^{151}$Raymond and Beverly Sackler School of Physics and Astronomy, Tel Aviv University, Tel Aviv; Israel.\\
% ...
$^{a}$ Also Affiliated with an institute covered by a cooperation agreement with CERN.\\
$^{b}$ Also at An-Najah National University, Nablus; Palestine.\\
$^{c}$ Also at APC, Universit\'e Paris Cit\'e, CNRS/IN2P3, Paris; France.\\
$^{d}$ Also at Borough of Manhattan Community College, City University of New York, New York NY; United States of America.\\
$^{e}$ Also at Center for High Energy Physics, Peking University; China.\\
$^{f}$ Also at Center for Interdisciplinary Research and Innovation (CIRI-AUTH), Thessaloniki ; Greece.\\
% ...

\end{flushleft}


\end{document}